\documentclass[11pt,aps,amsfont]{article}
\usepackage{bm}
\usepackage{feynmp}
\usepackage{appendix}
\usepackage{graphicx}
\usepackage{dsfont}
\usepackage{tipa}
\usepackage{a4wide}
\usepackage{amsmath}
\usepackage{graphicx}
\usepackage{amssymb}
\usepackage{amsbsy}
\usepackage{color}
\usepackage{dsfont}
\usepackage{upgreek}
\usepackage{stmaryrd}
\usepackage{hyperref}
\usepackage{bbm}
\usepackage{color}
\usepackage{setspace}
\usepackage{color}
\usepackage{amsmath}
\usepackage{amsfonts}
\usepackage{verbatim}
\usepackage{amssymb}
\usepackage{graphicx}
\usepackage{amsmath,amssymb,calc}
\usepackage[bulletsep]{collref}
\usepackage{bbm}
\usepackage{setspace}
\usepackage{color}
\usepackage{graphicx}
\usepackage{array}
\usepackage{caption}
\usepackage{subcaption}
\usepackage{adjustbox}
\usepackage{enumerate}
\hypersetup{colorlinks,%
    citecolor=blue,%
    filecolor=blue,%
    linkcolor=blue,%
    urlcolor=blue}
                               
\newcommand{\eqb}{\begin{eqnarray}}
\newcommand{\eqf}{\end{eqnarray}}
\newcommand{\bb}{\begin{equation}}
\newcommand{\ee}{\end{equation}}

\newcommand{\gp}{\gamma'}
\definecolor{gre}{rgb}{0,0.4,0.3}

\usepackage{color}

\newcommand{\muu}{m_{\gamma'}}
\newcommand{\mfi}{m_\phi}

\newcommand{\csq}{\cos^2\chi}
\newcommand{\ssq}{\sin^2\chi}
\newcommand{\cssq}{\cos^4\chi}
\newcommand{\T}{\text{T}}
\newcommand{\m}{\text{m}}
\newcommand{\eV}{\text{eV}}
\newcommand{\meV}{\text{meV}}
\newcommand{\csqa}{\cos^2\theta}
\newcommand{\ssqa}{\sin^2\theta}
\newcommand{\PLUS}{\left(\frac{\delta_1+\delta_2}{2}\right)}
\newcommand{\MINUS}{\left(\frac{\delta_1-\delta_2}{2}\right)}

\begin{document}
\date{}
\title{A two particle hidden sector and the oscillations with photons}
\author{
Pedro D. Alvarez$^1$,
Paola Arias$^2$\footnote{paola.arias.r at usach.cl},
Carlos Maldonado$^2$, \\[2ex] 
\small{\em $^1$Departamento de Fisica, Universidad de Antofagasta, Aptdo 02800, Chile} \\ 
\small{\em $^2$Departmento de F\'isica, Universidad de Santiago de Chile,} 
\small{\em Casilla 307, Santiago, Chile} \\[0.5ex]  
}
\maketitle

\begin{abstract}
We present a detailed study of the oscillations and optical properties for vacuum, in a model for the dark sector that contains axion-like particles and hidden photons. In this model, both can couple to photons. We provide bounds for the couplings versus the mass, using current results from ALPS-I and PVLAS. We also discuss the challenges for the detection of models with more than one hidden particle in light shining trough wall-like experiments.
\end{abstract}
\section{Introduction}
A  hidden sector weakly coupled to our Standard Model of particle physics has been so far the most straightforward way to introduce new physics into observable processes. Despite the success of both the actual Standard Model of particle physics and the cosmological $\Lambda$CDM model, there are still some awkward issues in high energy physics, such as dark matter, dark energy,  hierarchy, baryon asymmetry, among others.

A prime example where new physics has been invoked is the so called strong CP problem {\it ie.} the absence of processes that violate CP symmetry in the strong sector. The most elegant and accepted solution to this problem so far is the introduction of a new symmetry (the so-called Peccei-Quinn symmetry  \cite{Peccei:1977hh}) that it is spontaneously broken at a very high scale $f_\phi \gg v_{EW}$, higher than the electro-weak scale. As an unavoidable  heritage of the spontaneous symmetry breaking, a new pseudo-scalar boson emerges - named axion \cite{weinberg} -   naturally relaxing to zero CP violating term. The axion particle acquires an effective coupling to two photons, $\mathcal L_{\phi\gamma\gamma} \subset \frac{g_{\phi\gamma\gamma}}4 \phi F_{\mu\nu} \tilde F^{\mu\nu}$ which is key to search for such a particle \cite{Sikivie:1985yu}.  Soon enough, was realised that a hidden sector can feature such coupling, even when they do not solve the strong CP problem in string theory based extensions, therefore opening the door to the existence of the so-called axion-like particles, pseudo scalars that also can feature the two-photon vertex.   Secondly, the kinetic mixing term has become widely popular, since can be generically induced from string compactifications \cite{kinetic_mixing}. A new $U(1)$ gauge boson is kinetically mixed with the visible photon $\mathcal L_{\gamma \gamma'} \subset \chi F_{\mu\nu} X^{\mu\nu}$, where the dimensionless coupling $\chi$ it is predicted to be very small. The new gauge boson is typically dubbed hidden photon, or dark photon.   It is  assumed that these particles are  very light, since the corresponding symmetry groups are softly broken ({{no hair theorem: no global symmetries}}).

The search for axions, ALPs and hidden photons it is so far quite intense, specially because the three of them are viable cold dark matter candidates  \cite{Arias:2012az}. Both particle-scenarios have as a key feature, the oscillations  with the photon. This is of course  very welcomed,  and it is the main exploit in experimental setups to test these type of models. In the case of pseudo-scalars mixing with photons, an external electromagnetic field it is needed to make the oscillation possible, to conserve intrinsic angular momentum. Due to lack of positive signals, experiments are going into the realm of really high precision and intense sources. A whole new generation of upgrades is envisaged for the next years and several new ideas  have also emerged \cite{newideas}. In this sense, if any of the experimental searches finds a positive signal, it would really state a major breakthrough about the enlargement of the SM.

Given the importance and experimental efforts going into the direction to detect scalar and vector particles mixed with photons, seems perfectly valid to ask why not to consider a model where they both coexist, and even more, interact between themselves. Also, seems timely to take the opportunity to test new models of particle physics with the resources available in the present. 
In this context, we would like to explore a step further into the realm of light inhabitants of a hidden sector and consider a model with two hidden particles, which can interact with themselves, a pseudo-scalar boson plus an abelian U(1) gauge boson, also known as hidden photon. They both belong to the category of WISPs (Weakly Interacting Slim particles) if their mass is light. Has been already pointed out that such models can have a rich phenomenology, for instance in \cite{Masso:2005ym} they considered a model with a hidden photon coupled to an axion-like field, and the kinetic mixing term. Their aim was to find compatibility between the observation of the vacuum polarisation experiment PVLAS in 2007 \cite{pvlas} (discarded the following year) and astrophysical constraints.  Later in \cite{Jaeckel:2014qea}, a model with an axion-like particle + hidden photon was invoked to explain the 3.55 keV line in the spectra of galaxy clusters.  Another similar model has been recently considered \cite{Kaneta:2016wvf}, where the pseudo-scalar is the QCD axion, which is coupled to hidden photons via the term $g_{\gamma'\gamma'}\phi G_{\mu\nu}\tilde G^{\mu\nu}$ and the hidden photons are coupled to standard model photons via a kinetic mixing term $\chi F_{\mu\nu}G^{\mu\nu}$.

In this work, we have chosen to follow the construction from \cite{Jaeckel:2014qea}, therefore the hidden photon is the mediator between visible and hidden sectors. We are interested in the phenomenological consequences of the model, focusing - by now - on vacuum effects and finding the current sensitivity and expected sensitivity that available experiments have on this model . The manuscript is organised as follows: in section \ref{Lag} we write down the model we will work on, and establish the equations of motion for the mixing. In section \ref{prob} we find the probabilities of a photon oscillating into a WISP in a region where an homogeneous magnetic field is present. We also comment on the rotation and ellipticity that can acquire the photon source due to the mixing. We also write down useful approximations for the oscillation probabilities, classifying them according to the strength of the mixing between the ALP and the HP. In section {\ref{analysis}} we discuss the features of the model by first analysing ellipticity effects of a photon beam transversing a magnetic field region and secondly, a light-shaning-through-walls setup and we present  constraints to the parameters of the model. In section \ref{conclu}  we conclude. 
 
\section{The model}{\label{Lag}}
{Let us consider the effective Lagrangian for the model,
\eqb
\mathcal L=-\frac{1}{4}f_{\mu\nu}f^{\mu\nu} - \frac{1}{4}x_{\mu\nu}x^{\mu\nu} + \frac{1}{2}\partial_\mu\phi\partial^\mu\phi + \frac{1}{2}\sin\chi f_{\mu\nu}x^{\mu\nu} 
+ \frac{1}{4}g\phi x_{\mu\nu}\tilde{x}^{\mu\nu}  
- \frac{m_\phi^2}{2}\phi^2  
+ \frac{m^2_{\gamma'}\cos^2\chi }{2}x_\mu x^\mu,
\label{1}
\eqf
introduced in \cite{Jaeckel:2014qea}. In the equation above, $f_{\mu\nu}$ is the field strength associated with the photon ($a_\mu$) and $x_{\mu\nu}$ the counterpart of the hidden photon ($x_\mu$). Tilde symbolises their dual. We have assumed both ALP and hidden photon are massive. The coupling constant between ALP and hidden photon ($g$) is in general of the form $g\propto \alpha_x/(2\pi f_\phi)$, with $\alpha_x$ the analogue of the fine-structure constant in the hidden sector and $f_\phi$ the decay constant of the ALP. Let us note that only the hidden photon field is directly coupled to photons via the kinetic mixing term, parametrised by $\sin\chi$. Interactions between the ALP field and the photon will emerge via hidden photon mixing. This can be clearly seen if one performs the following redefinition 
\eqb
X_\mu&=& x_\mu - \sin \chi a_\mu\\
A_\mu&=& a_\mu \cos\chi.
\label{3}
\eqf
And the Lagrangian in this new basis reads
\eqb
\mathcal L'=&-&\frac{1}{4}F_{\mu\nu}F^{\mu\nu} - \frac{1}{4}X_{\mu\nu}X^{\mu\nu} + \frac{1}{2}\partial_\mu\phi\partial^\mu\phi + \frac{1}{4}g\phi X_{\mu\nu}\tilde{X}^{\mu\nu} +\frac{g}2 {\tan \chi }\phi X_{\mu\nu}\tilde{F}^{\mu\nu}\label{4} \\
 &&+\frac{g}{4}\tan^2\chi \phi F_{\mu\nu}\tilde{F}^{\mu\nu} 
- \frac{m^2_\phi}{2}\phi^2 + \frac{m^2_{\gp}\cos^2\chi}{2} \left(X_\mu X^{\mu} + 2\tan \chi X_\mu A^{\mu} + \tan^2\chi A_\mu A^{\mu}\right).\notag
\eqf
Note that in this new basis, the coupling between ALP-photon is manifest, with a coupling strength given by $g\tan^2\chi$. It is also manifest that a non-diagonal mass matrix appears, mixing photons and hidden photons. We now compute the equations of motion, obtaining
\eqb
-\partial_\mu F^{\mu\nu} +g \tan\chi \partial_\mu \phi \tilde{X}^{\mu\nu} + \tan^2\chi g\partial_\mu \phi \tilde{F}^{\mu\nu}&=&m_{\gp}^2\left( \cos\chi \sin \chi \,X^{\nu} +\sin^2 \chi\, A^{\nu}\right)
\label{5}\\
\partial_\mu \partial^{\mu}\phi + m_\phi^2\phi&=&\frac{g}{4} X_{\mu\nu}\tilde{X}^{\mu\nu} + \frac{g}{2}\tan\chi \left(X_{\mu\nu} + \frac{1}{2}\tan\chi F_{\mu\nu}\right) \tilde{F}^{\mu\nu}\\
-\partial_\mu X^{\mu\nu} + g\partial_\mu \phi \tilde{X}^{\mu\nu} +g\tan \chi \partial_\mu\phi \tilde{F}^{\mu\nu}&=&m_{\gp}^2\cos^2 \chi \left(X^{\nu} +\tan\chi \, A^{\nu}\right),
\label{7}
\eqf

\section{Vacuum Probabilities}{\label{prob}}
Let us start our analysis of the model with the simplest scenario, which would be the propagation of a photon beam in vacuum. For simplicity and following \cite{Raffelt:1987im} we will consider a coherent source of frequency $\omega$ propagating into the $\hat z$ direction. We assume the spatial extent of the photon beam  transverse to $\hat z$ is much bigger than the wavelength, and therefore the propagation can be thought of as one-dimensional. In order to have an interesting scenario (with oscillation between all three particles) we also assume the presence of an external homogeneous magnetic field $\vec B$, oriented in the $\hat x$ direction. We will also assume no external hidden field is present. We start by linearising the equations of motion, assuming that the external electromagnetic field is much stronger than the photon source $|\vec A_{\rm ext}| \gg |\vec A| $, and that terms of the form $\phi |\vec A |$, $|\vec A | |\vec X |$, $\phi  |\vec X |$ can be neglected. 

With these assumptions, the equations (\ref{5})-(\ref{7}) can be written as \footnote{We work in the gauge $\partial_i A_i=0$ and $A_0=X_0=0$.}
\begin{equation}
-({\partial_t}^2-\vec{\nabla}^2)\vec{A} -g \tan^2 \chi \,\partial_t \phi \vec{B}=m_{\gp}^2(\sin^2 \chi \,\vec{A} +\cos\chi \sin\chi  \, \vec{X})
\label{11}
\end{equation}
\begin{equation}
({\partial_t}^2-\vec{\nabla}^2)\phi + m_\phi^2=g\tan\chi \,\partial_t\vec{X}\cdot\vec{B} + g\tan^2\chi \,\partial_t\vec{A}\cdot\vec{B}
\label{12}
\end{equation}
\begin{equation}
-({\partial_t}^2-\vec{\nabla}^2)\vec{X} - g \tan\chi \partial_t \phi \vec{B}=m_{\gp}^2\cos^2\chi \left(\vec{X} + \tan \chi \,\vec{A}\right).
\label{13}
\end{equation} 

Since we are considering a beam source of photons with frequency $\omega$, propagating in the $\hat z$ direction, the plane wave approximation is a good one:
\begin{equation}
\vec{A}(z,t)=e^{i\omega t}\vec{A}(z),\,\,\,\,\,\,\,\, \phi (z,t)=e^{i\omega t}\phi (z), \,\,\,\,\,\,\,\,\, \vec{X}(z,t)=e^{i\omega t}\vec{X}(z)
\label{14}
\end{equation}

Replacing the ansatz in (\ref{11})-(\ref{13}) and taking the expansion{\footnote{Valid if the variation of the external field is negligible compared to photon wavelength in the considered scale and further the relativistic approximation $\omega+k\sim 2\omega$, where $k$ is the photon wave number \cite{Raffelt:1987im}. }} $(\omega^2+{\partial_z}^2)\approx{2\omega(\omega -i\partial_z)}$, we finally find:
\begin{equation}
\begin{pmatrix}
\omega -i\partial_z - \frac{m_{\gp}^2}{2\omega}
\begin{pmatrix}
 \sin^2\chi & \sin\chi \cos \chi \\
\sin\chi \cos \chi & \cos^2\chi
\end{pmatrix}
\end{pmatrix}
\begin{pmatrix}
A_{\perp} \\
X_{\bot}
\end{pmatrix}
=0,
\label{perpend}
\end{equation}
\begin{equation}
\begin{pmatrix}
\omega -i\partial_z - \frac{1}{2\omega}
\begin{pmatrix}
{m_{\gp}^2}\sin^2\chi & {m_{\gp}^2} \sin\chi \cos\chi &  {g B\omega\, \tan^2\chi }\\
{m_{\gp}^2\sin\chi \cos\chi}&{m_{\gp}^2}\cos^2\chi  & {g B\omega}\, \tan\chi  \\
{g B\omega\, \tan^2\chi } & {g B\omega\, \tan\chi } & {\mfi^2}
\end{pmatrix}
\end{pmatrix}
\begin{pmatrix}
A_{\parallel} \\
X_{\parallel}\\
\phi
\end{pmatrix}
=0.
\label{18}
\end{equation}
With the linearisation we obtain a first order differential equation of Schroedinger-type,  $i\partial_z \psi(z)=H\psi(z)$, where $\psi(z)=\left\{ A_\parallel (z), X_\parallel (z),\phi (z)\right\}$.
The symbols $\parallel$ and $\perp$ refer to the linear polarization of the photon parallel and perpendicular to the external magnetic field. 
As expected, the perpendicular component of the photon beam mixes only with the hidden photon, as in the standard scenario photon-HP. Instead, the parallel component  mixes with the ALP field and the  parallel polarization of the hidden photon. As we can see from eq.~({\ref{18}}), besides the usual couplings of $\phi$ and $X$ to the photon, there are couplings between ALP and HP, that will induce oscillations, in this linearised version triggered by the magnetic field.

To solve for the three fields amplitudes, we choose to introduce a $3\times 3$ rotation matrix, such that 
\eqb
\begin{pmatrix} A_\parallel \\
X_\parallel \\
\phi
\end{pmatrix}= R\begin{pmatrix} A'_\parallel \\
X'_\parallel \\
\phi'
\end{pmatrix} ; \,\,\,\,\,\,\,\,\,\,\,\,\,\,\, R^TR=I,\,\,\,\, {\mbox det}\,(R)^2=1.
\label{rot}
\eqf
$R$ corresponds to the matrix that diagonalises the Hamiltonian in equation ({\ref{18}}), $R^T HR=\text{diag}(\omega_1,\omega_2,\omega_3)$.  Therefore,  the primed states in the equation above are  propagation (mass) eigenstates, whose dispersion relation is given by
\eqb
\omega_1&=&\omega=k, \label{dispersion1}\\
\omega_2&=& \omega-\Omega{{-}}\Delta \label{dispersion2}\\
\omega_3&=&\omega-\Omega{{+}}\Delta.
\label{dispersion3}
\eqf
Where the functions $\Omega$ and $\Delta $ are defined respectively as
\begin{equation}
\Omega\equiv \frac{\muu^2+\mfi^2}{4\omega},\,\,\,\,\,\,\,\,\, \Delta\equiv \frac{gB}{2\cos^2\chi} \sqrt{\sin^2\chi+x^2 \cos^4 \chi}, \,\,\,\, \,\,\,\,\, x\equiv \frac{\muu^2-\mfi^2}{2 g B\omega}. \label{freqx}
\end{equation}

A convenient parametrisation of this matrix is given in terms of two angles, $\theta, \chi$ 
\begin{equation}
R=\begin{pmatrix} \cos\chi & \cos \theta \sin \chi & -\sin\theta \sin \chi\\
-\sin\chi & \cos\theta \cos\chi & -\cos \chi \sin \theta \\ 0 & \sin\theta  & \cos\theta.
\end{pmatrix}
\label{matrizR}
\end{equation}
Where,
\begin{equation}
\sin\theta= \frac{\sin\chi}{\sqrt{\mathcal F^2+\sin^2\chi}}, \,\,\,\,\,\,\,\,\,\,\,\,\, \mathcal F=\left(x +\frac{2\Delta}{Bg}\right) \cos^2\chi. \label{theta}
\end{equation}

The state $A_\parallel$ (and $A_\perp$, but we only focus on the parallel component for now) is firstly produced by the photon source, but since is not a mass eigenstate, it will oscillate into $X_\parallel$ and $\phi$, which are also not mass eigenstates.  Note that the states $X_\parallel $ and $\phi$ are sterile to matter currents. Using eq.~(\ref{rot}) and the matrix found in (\ref{matrizR}) we have (we will omit the parallel sub-index for now)
\eqb
A(z)&=& \cos \chi \,A'(z)+\sin \chi \,\Phi(z)\\
X(z) &=& -\sin \chi\, A'(z)+\cos \chi \,\Phi(z)\\
\phi(z)&= &\sin \theta \,X'(z)+ \cos\theta\,  \phi'(z).
\label{evolve0}
\eqf
Where we have taken $\Phi(z)= \cos\theta \, X'(z)-\sin\theta\, \phi'(z)$, and let us note that this state is orthogonal to $\phi(z)$. 
Primed fields are mass eigenstates, therefore their propagation can be found with the evolution operator $Y'(z)=U(z)Y'(0)$. 

In this basis, there are oscillations between $\Phi(z)$ and $\phi (z)$, which are parametrised and controlled  by the angle $\theta$. Their evolution  can be obtained from the evolution of $X'$ and $\phi'$, which have energies $\omega_2$ and $\omega_3$, respectively.  
\eqb
 \begin{pmatrix}
\phi(z)\\
\Phi(z) 
\end{pmatrix}= \begin{pmatrix} \cos\theta & \sin\theta \\ -\sin\theta & \cos\theta\end{pmatrix}
\begin{pmatrix} \phi'(z)\\ X'(z) 
 \end{pmatrix}.
 \label{evolve1}
 \eqf
 On the other hand, $\Phi$ can be also written as $\Phi= \cos\chi X+\sin \chi A$, therefore, the ALP field oscillates into this linear combination of photon and hidden photon and viceversa.
It is convenient to write the interaction of $A$ and $X$ in terms of the propagation state $A'$ and  the state $\Phi$, even though the latter it is not a mass eigenstate, its evolution in space is known, since gets  determined from the equation (\ref{evolve1}) above

\eqb
\begin{pmatrix}
A(z)\\
X(z) 
\end{pmatrix}= 
 \begin{pmatrix} \cos \chi &  \sin \chi \\  -\sin \chi & \cos \chi \end{pmatrix} \begin{pmatrix} A'(z)\\ \Phi(z) 
 \end{pmatrix}.
 \label{evolve2}
 \eqf
In this 'basis' the oscillation between $A$ and $X$ is controlled by the angle $\chi$. 
 It is now straightforward to find the amplitude of the fields at a distance $z$ from the origin, where we assume  as initial conditions, $A_\parallel (0)=1$ and $\phi(0)=X_\parallel(0)=0$:
\eqb
A_\parallel(z)&=& \cos^2\chi+ \sin^2\chi\left(e^{i(\Omega+\Delta) z} \cos^2\theta+e^{i(\Omega-\Delta)z} \sin^2\theta\right),\label{amplitudes}\\
X_\parallel (z)&=& {\sin\chi }\cos\chi\left(-1+ e^{i(\Omega+\Delta)z}\cos^2\theta+e^{i(\Omega-\Delta)z}\sin^2\theta\right),\notag\\
\phi(z)&=&-i\, {\sin\chi\, \sin 2\theta }\, \sin(\Delta z).\notag
\eqf
Where we have neglected a common phase $e^{-i\omega z}$ in all three amplitudes. 

Thus, the probability that after traveling a distance $z$ from the source, a photon will convert into an ALP, HP or remain a photon, is given respectively by
\begin{align}
 P_{\gamma_\parallel\rightarrow \gamma'_\parallel}=&4\csq \ssq \left(\csqa \sin^2\left(\frac{\delta_1+\delta_2}{2}\right)+\ssqa \sin^2\left(\frac{\delta_1-\delta_2}{2}\right)-\csqa\ssqa\sin^2 \delta_1\right)\,,\label{probabilities1}\\
 P_{\gamma\rightarrow \phi}=&4 \ssq \csqa\ssqa\sin^2 \delta_1\,.\label{probabilities2}
\end{align}
and $P_{\gamma_\parallel \rightarrow \gamma_\parallel}(z)=1-P_{\gamma_\parallel\rightarrow \gamma'_\parallel}-P_{\gamma_\parallel\rightarrow \phi}$. In (\ref{probabilities1}) and (\ref{probabilities2}) we have defined the oscillation angles by
\begin{equation}
 \delta_1=\Delta z\,, \qquad \delta_2=\Omega z\,.
\end{equation}

For the perpendicular component of the beam, the probability of oscillation to hidden photons it is known, but we write it by completeness
\bb
P_{\gamma_\perp\rightarrow {\gamma'}_\perp}=4\sin^2\chi \cos^2\chi \sin^2\left(\frac{\muu^2 z}{4\omega}\right).
\ee

We will analyse these probabilities below. Nevertheless, at this point it is important to mention that in this model - where we assume a negligible or zero background of hidden fields -   there will be a decoupling of the ALP field (from both HP and photons) for two different scenarios: when $\theta=\left\{0,n\pi\right\}$ and when $\theta=(2n+1)\pi/2  , \,n\in \mathds{Z}$. The first scenario corresponds to $g\times B \rightarrow 0$ and $\muu>\mfi$. The second one occurs also when $g\times B\rightarrow 0$, but $\mfi>\muu$. This can be seen explicitly from eqs.~(\ref{evolve0}), and from the definition of the angle $\theta$, eq.~(\ref{theta}).

\subsection{Rotation and ellipticity effects}
One of the first observable effects of the mixing studied in the previous subsection,  is the change in the original polarisation of the photon beam. In the case of the standard ALP-photon scenario, this effect is due to a uneven interaction of the ALP with the parallel and perpendicular components of the $A$ field with respect to the external magnetic field \cite{Maiani:1986md}. As a result of the mixing, the original wave  is rotated from the original polarisation plane and gets also a small elliptical component.  Experiments  dedicated to the vacuum magnetic linear birefringence, such as PVLAS \cite{pvlas} and BMV \cite{Cadene:2013bva}, have set constraints on the plane coupling of ALP-to-photon versus the mass of the scalar. In the scenario photon-hidden photon alone, there is no such effect, since the oscillation affects equally the perpendicular and parallel components of the photon field.

In our model these effects are also present. After transversing the magnetic region $z$, the beam has changed its amplitude and phase, as can be seen from eq.~(\ref{amplitudes}), meaning it develops a small elliptical component, and a rotation of the polarisation plane. Thus, the amplitudes evolve according to $A_\parallel (z)= a_\parallel(z) e^{-i\omega z+i\varphi_\parallel(z)}$ and $A_\perp (z)= a_\perp(z) e^{-i\omega z+i\varphi_\perp(z)}$. Since these are tiny effects, they can be written as $a_\parallel(z)=\cos(\alpha_0)(1-\epsilon_\parallel(z))$, $a_\perp(z)=\sin(\alpha_0)(1-\epsilon_\perp(z))$ and $e^{i\varphi_{\parallel(\perp)}}\approx 1+i\varphi_{\parallel(\perp)}$. Here $\alpha_0$ is the initial polarisation angle with respect to the direction of the magnetic field $\vec B$. The ellipticity and rotation angles  after traveling the magnetic region are given, respectively, by
\eqb
\psi=\frac{\sin(2\alpha_0)}2 \left(\varphi_\parallel -\varphi_\perp\right),\,\,\,\,\,\,\,\,\,\,\,\,\,\,\,
\delta \alpha=\frac{\sin(2\alpha_0)}2 \left(\epsilon_\parallel- \epsilon_\perp\right).
\eqf
After some manipulation of $A_\parallel (z)$ given by eq.~(\ref{amplitudes}) we find
\eqb
\varphi_\parallel&=&  \sin^2\chi \left[\sin\delta_2 \cos\delta_1+\cos 2\theta \sin\delta_1 \cos\delta_2\right],\\
\epsilon_\parallel&=& \sin^2\chi \left[1-\cos\delta_2\cos\delta_1+\cos2\theta \sin\delta_2 \sin\delta_1\right]. \label{paralelellip}
\eqf
While solving eqs.~(\ref{perpend}), we find
\eqb
\varphi_\perp&=&  \sin^2\chi \sin\left(\frac{\muu^2 z}{2\omega}\right) \label{perpendicular1}\\
\epsilon_\perp&=&2 \sin^2\chi \sin^2\left(\frac{\muu^2 z}{4\omega}\right).
\label{perpendicular2}
\eqf

In the next section we will find some interesting limits of these expressions, and compare it to the case of the ALP-photon scenario.  We will also use the above expressions to set constraints on the parameters of our model. 

\subsection{Limiting cases for the oscillation probability and resonances}

It is useful to analyse the behaviour of the mixing angle $\theta$ by computing $\tan2\theta$. From the definition of the mixing angle in (\ref{theta}) we have $\cos\theta = \mathcal F/\sqrt{\mathcal F^2+\sin^2\chi}$, then
\begin{equation}
\tan 2\theta= \frac{2\mathcal F \sin\chi}{\mathcal F^2-\sin^2\chi}.
\label{theta2}
\end{equation}

Having in mind the expression of $\mathcal F$, eq.~(\ref{theta}), we can recognise three different regimes:
\begin{enumerate}[(I)]
\item $|x|\ll \sin \chi \ll1$  
\item $\sin\chi \ll |x|\ll1$
\item $\sin\chi \ll 1 \ll |x|$.
\end{enumerate}

\vskip 2mm 
 {\it{  Strong mixing regime or resonance (I):}}\\
 \noindent In this limit, ${|\muu^2-\mfi^2|}/{2 g B\omega}\ll \chi\ll 1$,  therefore, the regime corresponds to very small masses of the WISPs respect to $gB$, or nearly degenerate  masses  $\muu\sim \mfi$. But we will analyse the degenerated case separately. We find the following approximations to be valid in this regime
{{ \begin{equation}
\Delta \approx \frac{gB}{2}\chi\left(1+\frac{x^2}{2 \chi^2}\right), \,\,\,\,\,\,\,\,\,\,\,\,  \mathcal F\approx x+ \chi
\end{equation} }}
Then, in the limit $x\ll \sin\chi\ll 1$ the angle $\theta$ gets determined according to
\begin{equation}
\tan 2\theta\approx \frac{\chi}x = \frac{2 \chi gB\omega}{\muu^2-\mfi^2}\gg1 .
\end{equation}
 Thus, the mixing between $\phi$ and the linear combination $\cos \chi \, X+\sin\chi\, A$ it is maximised. The angle $\theta \rightarrow \pi/4$ and it is given by
{{\begin{equation}
\theta_s\sim  \frac{\pi}4-\frac{\muu^2-\mfi^2}{4gB\omega \chi}.
\end{equation}}}
Above we have implicitly assumed $\muu>\mfi$. For the inverted case $x<0 $, meaning $\muu<\mfi$, we have $\theta\rightarrow -\pi/4-|x|/2\chi $.

Clearly, if the masses are very small respect to $gB$ (or nearly degenerated), then $\delta_1 \gg \delta_2$ in this regime. {Remarkably, even if both hidden particles are massless, the phenomenon of oscillation holds\footnote{But the magnetic field should still be on. As we have commented above if $B\rightarrow 0$, then $\theta \rightarrow 0$ $(\pi/2)$ when $m_{\gamma'}>m_\phi$ ($m_{\gamma'}<m_\phi$).}. This is contrary to what usually happens in vacuum in the simple models of photon-HP. The latter can be seen by explicitly writing the approximated expression for the oscillation probabilities in this regime, namely}
 {{\eqb
 & &P_{\gamma_\parallel\rightarrow\phi}(z)\approx \chi^2 \sin^2(\frac{gB\chi z} 2)\left[1-\left(\frac{\muu^2-\mfi^2}{2gB\omega \chi}\right)^2\right],\label{ALPstrong}\\
& &P_{\gamma_\parallel\rightarrow\gamma'_\parallel}(z)\approx 4\chi^2\sin^4(\frac{gB\chi z}4)+\chi \frac{\muu^2-\mfi^2}{gB\omega }\sin^2(\frac{gB\chi z}2).
\eqf

\vskip 1mm 
 {\it{  Weak mixing regime (II):}}\\
 \noindent This regime holds as long as $gB\chi \ll |\muu^2-\mfi^2|/(2\omega)\ll gB$, it is fulfilled. For this scenario, the relevant parameters of eq.~(\ref{freqx}), are given by
\eqb
&& \Delta \approx \frac{|\muu^2-\mfi^2|}{4\omega} +\frac{\chi^2 gB}{4|x|}, \label{deltaweak}\\
&& \mathcal F= \begin{cases} 2|x|+\frac{\chi^2}{2|x|}, \,\,\,\,\,\,\,\,\,\,\,\,\,\, {\rm{if}}\,\,\, x>0,\,\,\, |x|<1\\ \frac{\chi^2}{2|x|} \,\,\,\,\,\,\,\,\,\,\,\,\,\,\,\,\,\,\,\,\,\,\,\,\,\,\,\,\,\,\,\,\, {\rm if}\,\,\, x<0,\,\,\, |x|<1\end{cases}
\eqf 
Where the case $x>0$ corresponds to the hierarchy $\muu\gtrsim \mfi$ and $x<0$ to the inverted one $\muu\lesssim \mfi$ \footnote{It can also be that one particle is much more massive than the other.}. Let us first analyse the hierarchy $x>0$, the oscillation angle in this weak mixing regime, $\theta_w$, it is found to be
\eqb
&& \tan 2\theta_w \approx \frac{\chi}{|x|}+\mathcal O  \left(\frac{\chi^3}{|x|^3}\right)= \frac{2gB\omega \chi}{|\muu^2-\mfi^2|}\ll 1\\
&& \theta_w \approx  \frac{gB\omega \chi}{|\muu^2-\mfi^2|}.
\label{thetaweak}
\eqf
For this regime, the conversion of photons to ALPs it is suppressed compared to the hidden photon one, the probability of oscillation has the same shape as in the ordinary ALP-photon scenario, but suppressed as $\chi^2\theta_w^2$:
\eqb
P_{\gamma \rightarrow\phi} \approx 4\chi^2 \theta_w^2 \sin^2(\Delta L)=\frac{\chi^4 (g B L)^2}{4} \left|F(\delta_1)\right|^2, \,\,\,\,\,\,\,\,\,\,\,\,\,\, F(\delta_1)=\frac{1}{\delta_1} \,\sin(\delta_1).
\label{ALPweak}
\eqf
Where the function $F(\delta_1)$ is a form factor, which can be maximised such that $F(\delta_1)\rightarrow 1$, for $\delta_1\ll 1$.

The probability to convert into a hidden photon in this regime is
\eqb
P_{\gamma \rightarrow\gamma'}&&\approx 4 \chi^2\left(\sin^2(\frac{\delta_1+\delta_2}{2})-\frac{\theta_w^2}2\left(\cos(\delta_2-\delta_1)-\cos(\delta_1+\delta_2)\right)\right)-P_{\gamma\phi}.
\label{HPweak}
\eqf}
Where $P_{\gamma \phi}$ is the one obtained in (\ref{ALPweak}).

For the inverted hierarchy $\muu \lesssim \mfi$ (or $x<0$), the oscillation angle it is found to be
\eqb
&& \tan2\theta_{ inv} \approx -\frac{\chi}{|x|}+\mathcal O  \left(\frac{\chi^3}{|x|^3}\right),\\ 
&& \theta_{inv}\approx \frac{\pi}2-\theta_w.
\eqf
Where $\theta_w$ is the angle found in eq.~(\ref{thetaweak}). The probability of $\gamma-\phi$ oscillation is the same as found in (\ref{ALPweak}). While, the probability of oscillation between $\gamma-\gamma'$ it is found to be, 
\begin{equation}
P_{\gamma\rightarrow\gamma'}\approx 4\chi^2\left( \sin^2(\frac{\delta_2-\delta_1 }{2})+\frac{\theta_w}2 \left(\cos(\delta_2-\delta_1)-\cos(\delta_1+\delta_2)\right)\right)-P_{\gamma\phi}.
\end{equation}
Let us note that in this regime, ($x<0$), we have $(\delta_1+\delta_2)\sim \mfi^2 L/(2\omega)$ and $(\delta_2-\delta_1)\sim \muu^2L/(2\omega)$. Therefore, the leading oscillation frequency it is smaller than the inverted hierarchy case, eq~(\ref{HPweak}).

\vskip 2mm 
 {\it{ Very weak mixing regime (III):}}\\
 \noindent This regime holds for  ${\left| \muu^2-\mfi^2\right|}/{2gB\omega}\gg 1$ and weak kinetic mixing, $\chi \ll 1$. As in the previous case, we find the same limiting values for $\Delta$ and $\mathcal F$, and thus for the mixing angle $\theta$ :

\eqb
&& \Delta \approx \frac{|\muu^2-\mfi^2|}{4\omega} \sim \Omega, \label{deltaweak}\\
&& \mathcal \theta_w\approx \begin{cases} \frac{gB\omega \chi}{\muu^2-\mfi^2}, \,\,\,\,\,\,\,\,\,\,\,\,\,\,\,\, {\rm{if}}\,\,\, \muu>\mfi\\ \,\pi/2 \,\,\,\,\,\,\,\,\,\,\,\,\,\,\,\,\,\,\,\,\,\,\,\,\,\,\,\,\, {\rm if}\,\,\, \muu<\mfi, \end{cases}
\eqf 
Thus, for sufficiently weak mixing $|x|\gg 1$ there is a total decoupling of the ALP to the other particles. This is because the mixing of photon and HP is still mediated by $\chi$, but  $\theta$ in this regime it is even smaller than the weak mixing presented before. For masses of the ALP bigger than the HP we found $\theta=\pi/2$, and as we commented previously this is analogue to turning off the magnetic field. 

\vskip 2mm 
 {\it{ Special case of (I), degenerate masses: $\muu \sim m_\phi$}}\\
  \noindent In the case of degeneracy between  $\muu$ and $\mfi$, we are fully in resonance, meaning the mixing angle $\theta=\pi/4$. Also $x=0$, independently of the mass scale and $gB\omega$. Let us define $m\equiv m_\gamma \simeq\mfi$. Then $\Delta=gB\chi/2$ and $\Omega=m^2/2\omega$. Thus, we can recognise two different cases: when $\Delta>\Omega$ ($\delta_1>\delta_2$), or equivalently $\frac{m^2}{gB\omega}<\chi$,  strong field scenario. The  probabilities and polarisation parameters are  the same as the ones found in {\it i)}, replacing $\muu=\mfi=m$, therefore the expressions become mass independent. 

The second case is $\Delta<\Omega$ ($\delta_2>\delta_1$), or equivalently $\frac{m^2}{gB\omega}>\chi$. This is a weak field case. The conversion probability $P_{\gamma-\phi}$ is the same as the one found in (\ref{ALPstrong}). The conversion probability between photons and HPs it is changed though, since now the dominant frequency is $\Omega$:
\eqb
P_{\gamma\rightarrow\gamma'}&&= \chi^2\left(2-\cos(\delta_2-\delta_1)-\cos(\delta_1+\delta_2)-\sin^2 \delta_1\right)\\
&&\approx \chi^2\left( 4\sin^2(\frac{\delta_2}2)-\sin^2\delta_1\right).
\eqf

\subsection{Oscillation angles}

The oscillation angles $\delta_1$ and $\delta_2$ play a key role in the oscillation pattern, as can be seen from the probabilities in equation~(\ref{probabilities1}). Therefore, depending on the setup, one can find the optimal distance from the origin to place a photon detector in order to avoid minima of the conversion probability caused by the trigonometric dependence. In figure (\ref{fig:deltaomega}) we show the behaviour of these parameters as a function of $\mfi$, for two different masses $\muu$, and two configurations of $gB$. As can be seen, for very small masses of the WISPs, $\delta_1>\delta_2$. This regime is what we have called  '{\it{strong mixing regime}}'. As the masses of ALP and HP get bigger, eventually, they get of the same order $\delta_1 \sim \delta_2$.  
Therefore,  this dominance will depend, on the one hand on the competition between the scale  $|\muu^2-\mfi^2|/\omega$ and the strength of $g \times B$, which is parametrised by $|x|$ greater or smaller than one, see  eq.~(\ref{freqx}), and on the other hand between the competition of $x$ and the  kinetic mixing $\chi$.

\begin{figure}[t]
\begin{subfigure}{.5\textwidth}
\centering
\includegraphics[width=1\linewidth]{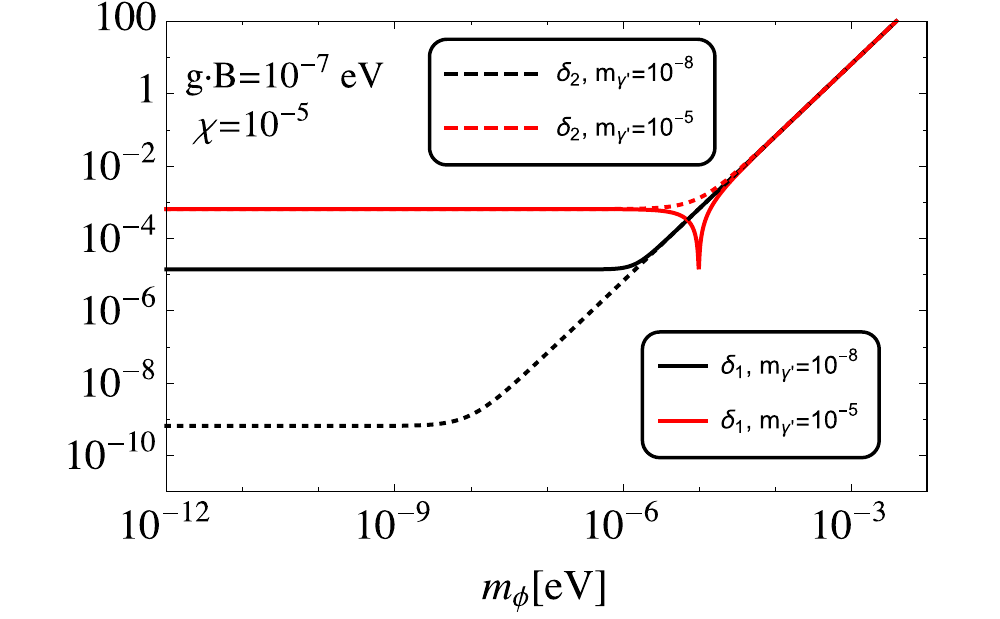}
\end{subfigure}
\begin{subfigure}{.5\textwidth}
\includegraphics[width=1\linewidth]{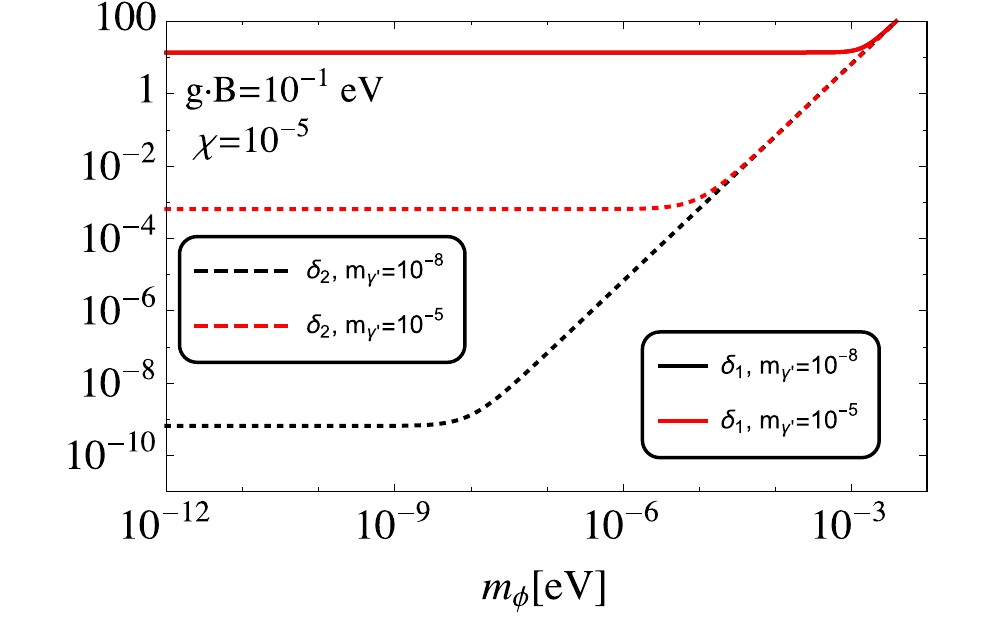}
\end{subfigure}
\caption{\small{The oscillation angles $\delta_1 =\Delta L$ and $\delta_2=\Omega L$ (defined in eq.~(\ref{freqx})) as a function of the ALP mass, $m_\phi$ for two different masses of HP. Plot of the left considers a 'small coupling' to the magnetic field, $g\times B=10^{-7} $~eV, while the plot of the right considers a 'strong coupling' with the magnetic field, $g\times B=10^{-1}$~eV. The magnitude of the angle $\delta_2$ is clearly dominated by the higher mass (between $\mfi$ and $\muu$), so for small masses we can have $\delta_1\gg \delta_2$ (this is the strong regime). As one of the masses gets bigger (always in comparison to $gB$), the oscillation angles become basically equal. This 'catching up' of $\delta_2$  to $\delta_1$ is highly dependent of the strength of $gB$.  }}
\label{fig:deltaomega}
\end{figure}

For benchmark values such as $B=5 \, \T$, $z\sim 10\, \m$ and $\omega=2.33\, \eV$, we get the following approximate expressions for the oscillation angles
\begin{equation}\label{delta1}
\delta_1 \approx\, 2\pi\left(\frac{z}{10\,\m}\right) 
     \begin{cases}
       \left(\frac{B}{5\,\T}\right)\left(\frac{g\chi}{3\times10^{-10}\,\eV^{-1}}\right)\left\{\frac{\sqrt{1+x^2 \cssq/\ssq}}{\csq}\right\},  & \text{if} \quad |x|<\sin \chi\\
       \left(\frac{2.33\, \eV}{\omega}\right)\frac{\muu^2-m_\phi^2}{(10^{-3}\, \eV)^2}\left\{\sqrt{1+\ssq/(x^2 \cssq)}\right\},  & \text{if} \quad |x|>\sin \chi\\ 
     \end{cases}
\end{equation}
\begin{equation}\label{delta2}
 \delta_2\approx\, 2\pi \left(\frac{z}{10\,\m}\right) \left(\frac{2.33\, \eV}{\omega}\right)\frac{\muu^2+m_\phi^2}{(10^{-3}\, \eV)^2}
\end{equation}

where the coefficients between brackets $\{ \cdot \}$ are only relevant when $|x|\approx \sin \chi$, but can be approximated as $\{ \cdot \}\approx 1$ otherwise.

By looking at the expressions (\ref{delta1}) we can see that we are in the small $\delta_1$-angle regime for sufficiently small couplings, $g \sin \chi < 3\times 10^{-10}\, \eV^{-1}$, in case (I) ($|x|<\sin \chi$) or when $|\muu^2-m_\phi^2|<(10^{-3}\,\eV)^2$, in case (II) or (III) ($|x|>\sin \chi$). The latter occurs when either both masses are smaller than $10^{-3}\,\eV$ or when we have fine tuning of the masses in such a way that the difference is smaller than $10^{-3}\,\eV$ in absolute value.

From (\ref{delta2}) we see that the small $\delta_2$-angle regime occurs for $|\muu^2+m_\phi^2|<(10^{-3}\,\eV)^2$, so both masses have to be smaller than the $\,\meV$.

In the small-angles regime, the baseline of the chamber is not long enough to have one complete oscillation and for this reason we do not need to worry about conversion probabilities arriving at a minima at the detector.

Let us now determine the range of parameters where one of the angles dominates, and when the oscillation angles are comparable, in which case there could appear suppressions in the conversion probability (\ref{probabilities1}). In case (I) we have
\begin{equation}
 \delta_2/\delta_1\approx \frac{\muu^2+m_\phi^2}{|\muu^2-m_\phi^2|}\times\frac{|x|}{\sin \chi}\,,
\end{equation}
which tell us that $\delta_2/\delta_1<1$, except if the masses are tuned to make the the denominator small enough. In order to get $\delta_2/\delta_1>1$ we need a fine tuning $\xi$, of at least $\xi<|x|/\sin \chi$, where $m_\phi=\muu(1\pm\xi)$. We can also have $\delta_2\approx\delta_1$ when $\xi\approx|x|/\sin \chi$, and this will suppress the second term in the conversion probability (\ref{probabilities1}). In case (I), however, the mixing angle $\theta$ is close to $\pi/4$ so the first term in (\ref{probabilities1}) is not neglectable.

In cases (II) and (III) we will have
\begin{equation}
 \delta_2/\delta_1\approx \frac{\muu^2+m_\phi^2}{\muu^2-m_\phi^2}\,,
\end{equation}
so generically $|\delta_2/\delta_1|>1$, except when one of the masses is much bigger than the other one in which case we will have $\delta_2\approx \delta_1$ for $\muu\gg m_\phi$ or $\delta_2\approx -\delta_1$ for $\muu\ll m_\phi$. Fortunately, the first two terms in (\ref{probabilities1}) cannot be suppresed simultaneously because when $m_{\gamma'}>m_\phi$ ($m_{\gamma'}<m_\phi$) then $\theta\sim 0$ ($\theta\sim \pi/2$).

\vskip 0.5cm

\section{Analysis of observable effects}{\label{analysis}}
\subsection{Ellipticity angle in vacuum-birefringence-type experiments  }
As mentioned in previous section, a feature of these kinds of models is the change in rotation and ellipticity angle of a polarised beam of photons. We will not analyse the rotation effects, since in the next section we make a detailed analysis of LSW experiments, which feature many of the effects we also find in the rotation of the polarisation plane.
\begin{figure}
\center
\includegraphics[scale=0.7]{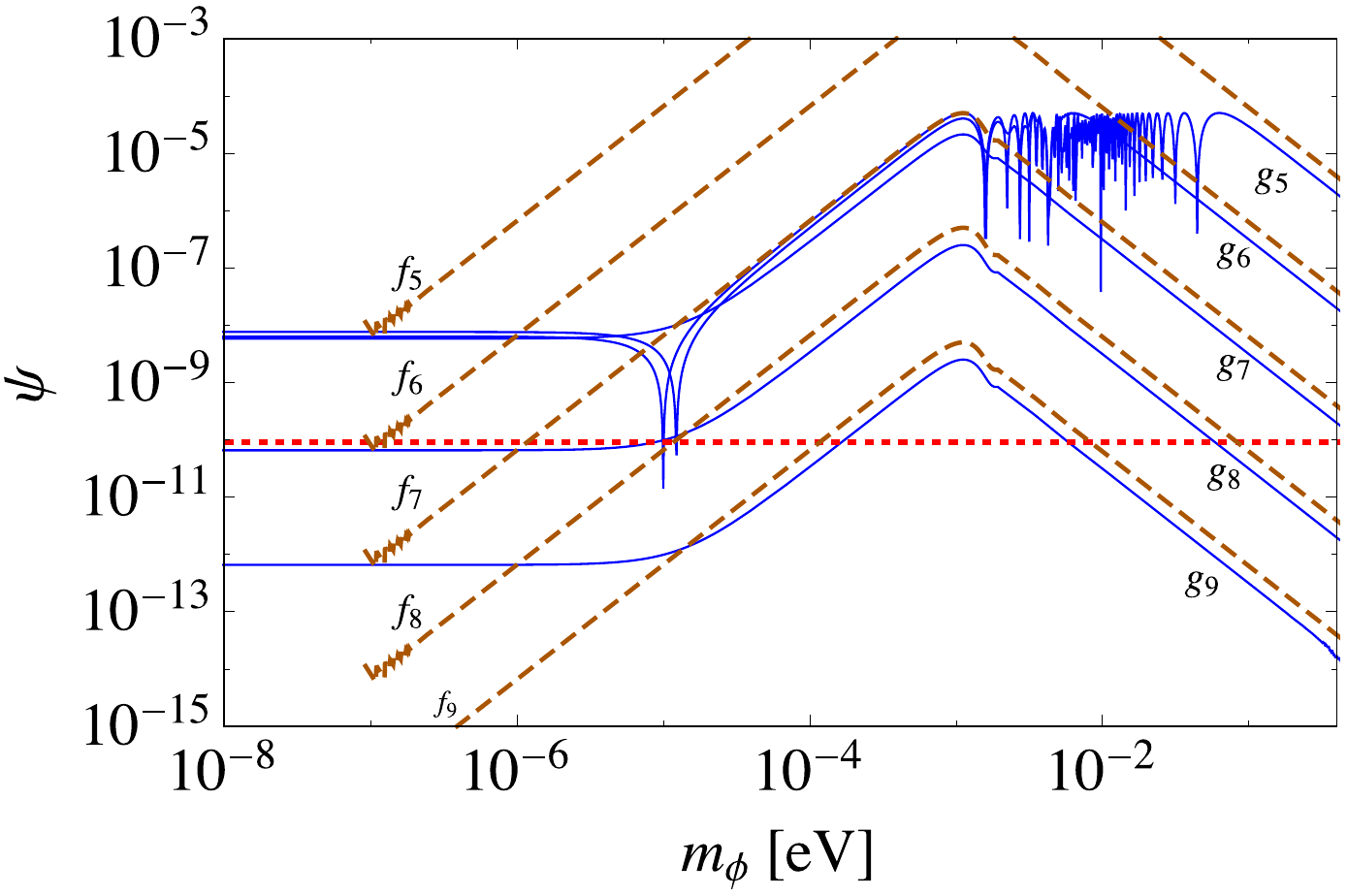}
\caption{\small{Comparison of ellipticity effects in the model of ALP-photon (dashed) and the ALP-HP-photon model (solid line). The nomenclature is $f_i\rightarrow g_{\phi\gamma\gamma}=10^{-i}$ GeV$^{-1}$ and $g_i \rightarrow g=10^{-i}$ eV$^{-1}$. The benchmark values used are $\muu=10^{-6}$~eV, $B=2.6$~T, $\chi=10^{-2}$, $\omega=1$~eV, $L=1$~m. Red dotted line corresponds to the experimentally measured value of the ellipticity angle from PVLAS \cite{pvlas}, $|\psi_{exp}|\leq 9\times 10^{-11} $.}}
\label{elipti}
\end{figure}

Changes in ellipticity angle of the photon beam after traveling through a region with magnetic field can be a very distinctive feature of the present model. For instance, in figure (\ref{elipti}) we show a comparison between the change in ellipticity angle due to the common  ALP-photon model (dashed dark orange) compared to the model analysed here (blue line).  As can be seen, there is a region in parameter space where both expressions coincide, making the identification $g_{\phi\gamma\gamma}\rightarrow g\chi^2$, with $g_{\phi\gamma\gamma}$ the coupling of the pure ALP-photon model. To understand better the behaviour of the ellipticity changes, let us look at the expression for the ellipticity angle $\psi$, defined by eq.(\ref{paralelellip}):
\begin{equation}
\psi \propto\varphi_\parallel-\varphi_\perp= \sin(\chi^2)\left(\sin\delta_2\,\cos\delta_1+\cos2\theta\, \sin\delta_1\,\cos\delta_2\right)-\sin\chi^2\, \sin(\frac{\muu^2 L}{2\omega}).
\label{elip}
\end{equation}
Where the last term corresponds to the contribution to the ellipticity angle from the perpendicular component of the beam, caused by the mixing of the photon with the HP alone.  Let us first consider the parameter space where $|x|< \chi$, {\it i.e.} the strong mixing regime. 

Let us consider $\delta_1,\delta_2\ll1$. For this parameter space $\theta\sim \pi/4-(\muu^2-\mfi^2)/(4gB\omega \chi)$, thus replacing in the above equation we find
\bb
\psi \approx \chi^2 \left( \frac{\muu^2+\mfi^2}{4\omega}L\, \cos\delta_1+\frac{\muu^2-\mfi^2}{4\omega}L \,\cos\delta_2- \frac{\muu^2 L}{2\omega}\right)\approx 0.
\label{cancel1}
\ee
Thus, when $\delta_1\ll 1$ and $\delta_2\ll1$, both expressions cancel and in this parameter space it is not possible to constrain the model. 
We can make a rough estimation about  the values of $g$ that can not be tested due to this cancelation from $\delta_1\ll1$, we obtain
\bb
g \lesssim 10^{-7} \mbox{eV}^{-1} \left(\frac{1 \mbox{T}}{B}\,\frac{10^{-2}}{\chi}\,\frac{\mbox{1\,m}}{L}\right)
\ee
On the other hand - also in the strong mixing regime - for $\delta_1 \gg 1$ and $\delta_2\ll1$, eq.~(\ref{elip}) can be approximated as
\bb
\psi \approx \chi^2 \left( \frac{\muu^2+\mfi^2}{4\omega}L \,\cos \delta_1- \frac{\muu^2 L}{2\omega} \right).
\label{elip_gaps}
\ee
Again, when $\muu\geq \mfi$, the expression can be simplified further to
\bb
\psi \approx \chi^2  \frac{\muu^2}{4\omega}L \left(\cos\delta_1 -2\right).
\label{eqa}
\ee
The above expression can not vanish, the smaller value it can take is when $ \delta_1 = 2n\pi$, and the largest (in absolute value), when $\delta_1= (2n+1)\pi$, with  $n \in Z$. In such case the maximum ($+$) and minimum ($-$) values for the ellipticity angle are approximated as 
\bb
\psi_{-} \approx -\chi^2  \frac{\muu^2}{8\omega}L,\,\,\,\,\,\,\, \psi_{+}=-3\chi^2  \frac{\muu^2}{8\omega}L,
\ee
where we have assumed an initial polarisation angle of the beam of  $\alpha_0=45^o$. Thus in this regime the ellipticity angle it is independent of the mass of the ALP. If $|\psi_-| \leq |\psi_{\rm{exp}} |\leq |\psi_+|$ -  where $\psi_{\rm{exp}}$ corresponds to the experimental measured value of the ellipticity angle- we expect to observe sensitivity gaps on the plane $g$ vs. $m_\phi$, as can be seen in figure ({\ref{fig:ejemplo_elip}}). This gaps will be placed in
\bb
g=\frac{4\pi n}{B\chi L}, \,\,\,\,\,\, \Rightarrow \,\,\, g= 5.1\times n \times 10^{-7}\, \mbox{eV}^{-1}\, \left(\frac{2.5 \rm{T}}{B}\, \frac{10^{-2}}{\chi}\, \frac{1\rm{m}}{L}\right), \,\,\,\,\, n\in Z.
\label{ggaps}
\ee
The gaps in sensitivity should, in principle, not be a problem experimentally, since they can be removed by changing the magnetic field (or the length of the experiment).  Finally, if $|\psi_{\rm{exp}}| \lesssim |\psi_-|$, then the parameter space gets covered completely, with no gaps.  And for the opposite  $|\psi_{\rm{exp}}| \gtrsim |\psi_+|$, the parameter space gets unconstrained.

Let us now move on,  if the parameters satisfy $\chi<|x|$ (weak mixing regime), eq.~(\ref{elip}) it is better approximated by
\begin{equation}
\psi \propto \chi^2\left(\sin(\delta_2\pm\delta_1)\mp2\left(\frac{g \chi B\omega}{(\muu^2-\mfi^2)}\right)^2 \sin \delta_1\, \cos\delta_2 -\sin(\frac{\muu^2}{2\omega}L)    \right)
\end{equation}
Where the upper sign accounts for $\muu>\mfi$ and the lower sign, for $\muu<\mfi$, respectively. If $|x|\gg \chi$, and $\muu<\mfi$, we have $\delta_2-\delta_1\approx \muu^2L/(2\omega)-(\chi g B)^2\omega L/(2\mfi^2)$, and the expression can be  approximated as
\begin{equation}
\psi \propto\chi^2\frac{g^2B^2\omega^2}{\mfi^4}\left(-\frac{\mfi^2 L}{2\omega}+ \sin(\frac{\mfi^2 L}{2\omega})\right),
\label{aproxvar}
\end{equation}
where we have assumed $(\chi g B)^2\omega L/(2\mfi^2)\ll 1$, which is a reasonable assumption when $|x|\gg\chi$. We note the equation above has the same form of the ellipticity change induced by axion-like particles \cite{Raffelt:1987im} in the weak mixing regime, namely
\begin{equation}
\psi_{ \text{alp}}=\frac{g_{\phi\gamma\gamma}^2B^2\omega^2}{\mfi^4}\left(y-\sin y\right), \,\,\,\,\,\,\ y=\frac{\mfi^2 L}{2\omega},
\end{equation}
if the replacement $g_{\phi\gamma\gamma}\rightarrow g\chi^2$ it is made. This behaviour it is shown in figure~(\ref{elipti}) where we compare the ellipticity angle discussed in this model, to the ellipticity angle from the ALP-photon model alone. We can see that in the region of $\mfi\gtrsim 10^{-4}$~eV, both expressions seem to coincide up to a saturation regime, where $g$ is already too big to fulfil the condition $|x|>1$, and therefore eq.~(\ref{aproxvar}) no longer holds. Let us note that the above expression does not depend on the HP mass. 

For the inverted hierarchy $\muu>\mfi$ and $|x|\gg \chi$ we find the approximated expression
\begin{equation}
\psi \propto \chi^2\frac{(g B \omega)^2}{\muu^4}\left(q-\sin q\,\left(1+\frac{\muu^4}{(g B \omega)^2}\right)\right),\,\,\,\,\,\,\,\,\, q=\frac{\muu^2 L}{2\omega}.
\end{equation}
Therefore, the shape of the ellipticity angle remains almost the same, but the oscillation angle depends on the mass of the HP. 
\begin{figure}[t]
\begin{subfigure}{.5\textwidth}
{\includegraphics[scale=0.5]{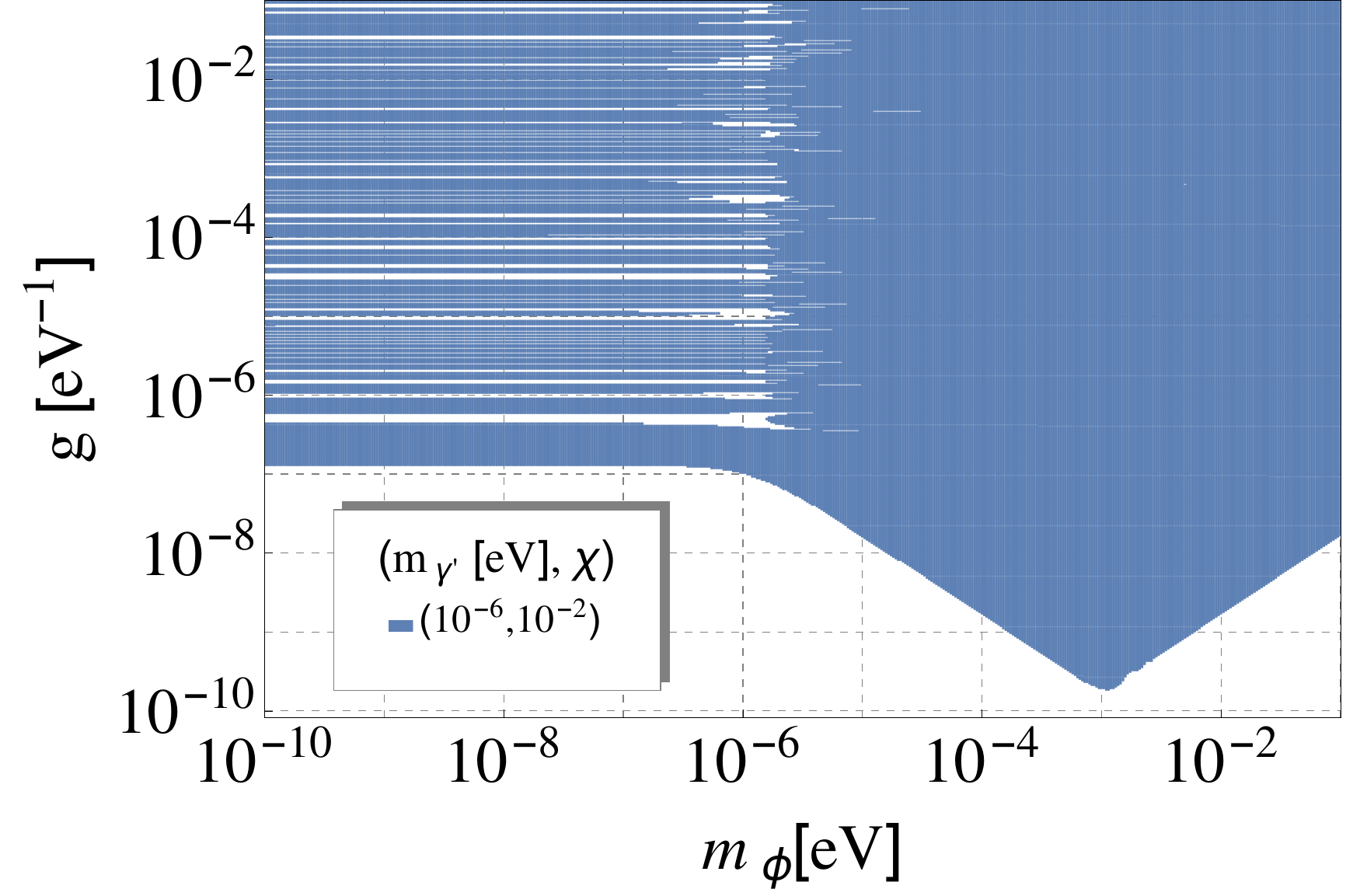}}
\end{subfigure}
\begin{subfigure}{.5\textwidth}
{\includegraphics[scale=0.5]{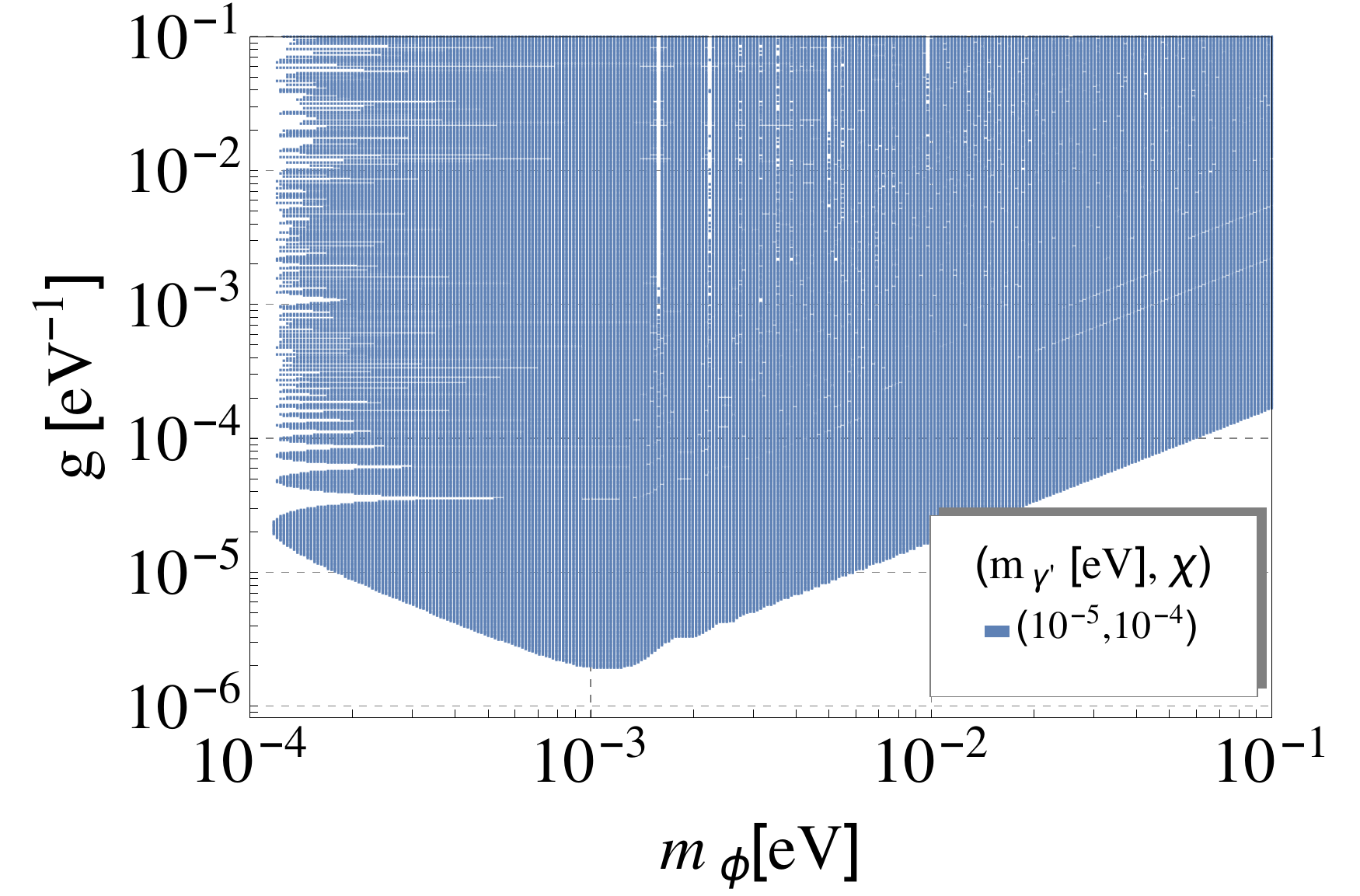}}
\end{subfigure}
\caption{\small{Exclusion plot for the ALP using ellipticity measurements. We have taken $\psi_{\rm exp}=9\times 10^{-11}$ as suggested by the results of PVLAS \cite{DellaValle:2015xxa}. Left figure: For small ALP masses the parameter $g$ is independent of the mass (strong regime) but some gaps in sensitivity appear because the oscillation angle $\delta_1>1$ for this parameter space. For $\muu<\mfi$ the HP has almost no effect on the ellipticity angle, so the ALP gets more constrained. Right figure: the smallness of $\chi$ prevents for being sensitive to small ALP masses. When $\mfi\sim 10^{-4}$, we have  $\chi^2 \mfi^2 L/\omega \sim \psi_{\rm exp}$ and the experiment can have again some sensitivity to the ALP parameter space. For both figures  we have considered the benchmark values $B=2.5 $~T, $L=1$~m and $\omega =1~$eV. }}
\label{fig:ejemplo_elip}
\end{figure}

In figure(\ref{fig:ejemplo_elip}) we show two examples for the exclusion in the $g$ vs.~$\mfi$ parameter space using ellipticity measurements that feature the behaviour previously analised. On the l.h.s. it is not possible to constrain the parameter space $g\lesssim 10^{-7}$~eV$^{-1}$ for small $\mfi$ masses because of the cancellation of eq.~(\ref{cancel1}). As $g$ gets bigger, $\delta_1\sim 1$ and eq.~(\ref{elip_gaps}) holds. For the chosen benchmark values of the experiment, (see caption for details), the parameter space $g\gtrsim 10^{-7}$~eV$^{-1}$ and $\mfi<\muu$, fulfils the criteria for the appearance of gaps in sensitivity, positioned in eq.~(\ref{ggaps}). Finally, for masses $\mfi>\muu=10^{-6}$~eV, the parameter space fits the weak mixing regime, so eq.~(\ref{aproxvar}) holds, and we see the characteristic  "V" behaviour for the sensitivity of the ALP, also appearing for the ALP-photon model, see fig.~(\ref{elipti}).

Meanwhile, in the r.h.s. of figure~(\ref{fig:ejemplo_elip}) we have considered a HP of $\muu=10^{-5}$~eV and $\chi=10^{-4}$. For these parameters it is not possible to have sensitivity to $\mfi\lesssim 10^{-4}$~eV, since $\psi_{\rm exp} \sim  9\times 10^{-11}>\varphi_{\perp,\parallel}$ for those masses. There are again some gaps in sensitivity due to the cancellation of the trigonometric functions. But these gaps can be covered experimentally by changing $B$ and or $L,\omega$.
\subsection{A light-shining-through-walls scenario}
Another very interesting scenario to test the oscillation effects are the Light-shining-through-walls (LSW) experiments. They consist of two identical magnetic regions of length $L$, separated by an opaque wall. In the first region, photons from an intense source (such as a $\approx$ 30W laser) are shone to the wall. Photons can oscillate into WISP and these can pass through the wall, and be reconverted to photons in the second magnetic region.  LSW experiments are currently under major upgrades, for instance, the ALPS experiment at DESY \cite{Spector:2016vwo}. Among the most important upgrades, are the length of the magnetic region, the use of frequency-locked Fabry-Perot cavities and single-count-photon detectors. 

In pure HP-photon model, oscillations occur thanks to a mixing via a non-diagonal mass matrix. However, in the presently studied model, there are also oscillations between HPs and ALPs, triggered  by the presence of a magnetic field. Therefore, seems possible that LSW experiments could lose some sensitivity due to the possibility that oscillations among WISPs are favoured, thus diminishing the probability to regenerate a photon in the second cavity. 
To compute this probability we first start assuming that the photon gets absorbed by the wall, so $A(L+\delta L)=0$, where $\delta L$ is the thickness of the wall and $L\gg \delta L$.  Neglecting the thickness of the wall compared to the length of the magnetic region, the propagation states $A',X', \phi'$ and the interaction state $\Phi$ right after the wall are found to be
\eqb
 A'(L+\delta L)&=&  \phi(L)\, \sin\theta+ \Phi(L)\, \cos\theta,\\
 \phi'(L+\delta L)&=& \phi(L)\, \cos\theta-\Phi(L)\, \sin\theta,\\
 \Phi(L+\delta L)&=& X (L)\, \cos\chi.
\eqf
Now it is  straightforward to evolve the propagation states in space and find the amplitude of photons, HPs and ALPs at $z=2L$, using eqs.~(\ref{evolve1}) and (\ref{evolve2}). We find the photon amplitude to be
\eqb
A_{\parallel}(2L)=\sin ^2\chi  \left(\cos ^2\chi  \left(\cos ^2\theta +\sin
   ^2\theta e^{-2 i \delta_1}-e^{-i  (\delta_1 +\delta_2
   )}\right)^2+\sin ^2\theta  \cos ^2\theta  \left(-1+e^{-2
   i \delta_1}\right)^2\right)\,.
\eqf
Where we have omitted a common phase of $ e^{2i(\delta_1+\delta_2)}$. It can be easily check that $P_{\gamma_\parallel\rightarrow\mathcal W \rightarrow\gamma_\parallel}=|P_{\gamma_\parallel\rightarrow\phi}(L)+P_{\gamma_\parallel\rightarrow\gamma'_\parallel}(L)|^2$, where $P_{\gamma_\parallel\rightarrow\phi}$ and $P_{\gamma_\parallel\rightarrow\gamma'_\parallel}$ are given by (\ref{probabilities1}) and (\ref{probabilities2}) respectively, the explicit expression is given by
\begin{equation}
 P_{\gamma_\parallel\rightarrow\mathcal W \rightarrow\gamma_\parallel}=16 \sin^4\chi \left(\cos^2\chi\left(\sin^2\theta\sin^2\MINUS+\cos^2\theta\sin^2\PLUS \right)+\sin^2\chi\sin^2\theta\cos^2\theta\sin^2\delta_1 \right)^2\,,
\end{equation}

The perpendicular component, on the other hand $P_{\gamma_\perp\rightarrow\mathcal W \rightarrow\gamma_\perp}=P_{\gamma_\perp\rightarrow\gamma'_\perp}(L)^2$, and it is given by
\begin{equation}
 P_{\gamma_\perp\rightarrow\gamma'_\perp}(L)=16 \sin^4 \chi \cos^4 \chi \sin^4\left(\frac{\muu^2 L}{4\omega} \right)
\end{equation}

The total probability for photon regeneration has to include both parallel and perpendicular components with appropriate weights in order to take into account the polarization of the beam,
\begin{equation}
 P_\text{LSW}= w_\parallel P_{\gamma_\parallel\rightarrow\mathcal W \rightarrow\gamma_\parallel}+w_\perp P_{\gamma_\perp\rightarrow\mathcal W \rightarrow\gamma_\perp}\,.
\end{equation}

We will again write approximate expressions for $P_{\gamma_\parallel\rightarrow\mathcal W \rightarrow\gamma_\parallel}$ that depend on the regime of the parameters and discuss the physics briefly.

\vskip 2mm
{\it{Case (I), $|x|\ll \sin\chi$:}}\\
\noindent In the case of non-degenerate resonance, the probability reads,
\begin{equation}
P_{\gamma_\parallel\rightarrow\mathcal W \rightarrow\gamma_\parallel}\approx4 \chi^4 (1-\cos \delta_1 \cos \delta_2)^2\,,
\label{approxres}
\end{equation}
Therefore, in this regime the probability is insensitive to the mass of the WISPs except for when the oscillation angle $\delta_2$ is out of the small-angle approximation. Recall that the latter only occurs for $|m_{\gamma'}^2+m_\phi^2|>\meV$, which is a small subregion of the space of parameters. In the pure photon-ALP scenario $P_{\gamma_\parallel\rightarrow\phi \rightarrow\gamma_\parallel}$ is also mass-independent and that is because the mixing is due to the presence of the $B$ field. However, for the pure HP-photon model the coupling among them vanishes as the mass of the HP gets smaller and so does vanish $P_{\gamma_\parallel\rightarrow\gamma'}$.
 
The approximate formula for the probability in this regime allow us to get an idea of some features  of the model. From (\ref{approxres}) we see that that the oscillation can  indeed be suppressed when either one of the following conditions holds
\begin{equation}\label{cond1}
 \delta_1\approx2 n \pi  \quad \text{and} \quad \delta_2\approx2 m \pi \,, \quad m,n \in  \mathbb{Z}\,,
\end{equation}
or 
\begin{equation}\label{cond2}
 \delta_1\approx(2 n+1) \pi  \quad \text{and} \quad \delta_2\approx(2 m+1) \pi \,, \quad m,n \in  \mathbb{Z}\,.
\end{equation}

Let us discuss the location and shape of this regions in a exclusion plot $\chi$ vs $m_{\gamma'}$ or $g$ vs $m_\phi$.

For small masses, $|m_{\gamma'}^2+m_\phi^2|<\meV$, we have $\delta_2< 1$ and in the generic case (in the sense of no tuning of the masses) we have $\delta_1 > \delta_2$, that means that only the first condition (\ref{cond1}) can holds for $m=0$ and $n\in\mathbb{Z}$. This will produce narrow fringes of insensitivity around the lines $\delta_1=2 n \pi$ or
\begin{equation}
g\chi=\frac{4\pi n}{BL}, \quad n \in \mathbb{Z}.
\end{equation}
These fringes correspond to the horizontal lines in plots of figures (\ref{fig:alps1,(mphi,g)-plane}) and (\ref{fig:alps1,(mx,chi)-plane}) that clearly can only occur in regions where $g\chi > 3\times 10^{-10} \eV^{-1}$ (using ALPS-I fiducial values for $B$ and $L$). Unfortunately, the approximate formula cannot capture all the details of the shape and thickness of the stripes. In figure (\ref{fig:alps1,(mphi,g)-plane}) we can see that for bigger values of $\chi$ the gaps in sensitivity become very narrow or completely disappear, which is expected for bigger couplings. These gaps of sensitivity can be removed by changing slightly the position of the detector or the magnitude of the magnetic field when $n \gg 1$.

The approximate formula also predicts that if the masses are small and tuned (in the sense defined above in the text with $\xi < |x|/\sin \chi <<1$), we have $\delta_2 > \delta_1$, which means that only one line of insensitivity survives when $m,n=0$ or $g\chi=0$, if $\delta_1\ll 1$. This, however, does not occur when the full formula is used and the gaps in sensitivity remain there even if the masses are tuned.

For bigger masses, $|m_{\gamma'}^2+m_\phi^2|\sim\meV - 10^{-2} \eV$ (but small enough to still be in case (I), see region near to but above the dashed line in fig. \ref{fig:alps1,(mphi,g)-plane}), both conditions (\ref{cond1}) or (\ref{cond2}) can hold for certain $m$, $n$ that are sufficiently big. Assuming $\chi$ is big enough, this will produce rounded regions of insensitivity around the points given by (\ref{cond1}) and (\ref{cond2}) when the equality holds (those are points in the ($\chi$,$m_{\gamma'}$) or ($g$,$m_\phi$) planes). If there is enough overlapping between the rounded regions they will merge and look like stripes in diagonal, see right hand side of plots in figure (\ref{fig:alps1,(mphi,g)-plane}) for small enough coupling $\chi$. These gaps in sensitivity can be removed by changing $B$ or $L$ slightly, when $n\gg 1$, or by changing the refraction index (by using a gas with the right density in the chamber) when $n$ is a small integer.

\vskip 2mm
{\it{Cases (II) and (III), $\sin \chi \ll |x|$:}}\\
\vskip 2mm
In this case the approximate formula simplifies to
\begin{equation}\label{PapproxII&III}
 P_{\gamma_\parallel\rightarrow\mathcal W \rightarrow\gamma_\parallel} \approx
 16 \chi^4 \sin^4 \left(\frac{\muu^2 L}{4\omega}\right)+O(\chi^6/x^2)\,.
\end{equation}
There is effectively a single oscillation angle that depends on the mass of the hidden photon only because the oscillation angles combine as $\frac{\delta_1\pm \delta_2}{2} = \frac{\muu^2 L}{4\omega}$ for $x>0$ ($x<0$) for the upper (resp. lower) sign.

We can have suppression in the probability when $\frac{\muu^2 L}{4\omega}\approx \pi n$ ($n$ an integer), in which case we need to consider the subleading term
\begin{equation}
O(\chi^6/x^2)=\frac {8\chi^6} {x^2}\sin \delta_1 \sin^2 \left(\frac{\muu^2 L}{4\omega}\right) \left(\mp\sin \delta_2+ \sin \delta_1 \chi^2 \right)\,.
      \end{equation}
The subleading term would restore a dependence in $m_\phi$ of the probability, as can be seen in plots of figure (\ref{fig:alps1,(mphi,g)-plane}) below the dashed line.

The oscillations of (\ref{PapproxII&III}) can be seen as oscillations of the black line in the plots of figure (\ref{fig:alps1,(mx,chi)-plane}) because when $B=0$ the subleading term vanishes. When $B$ is turned on, some of these gaps in sensibility are filled up, leaving behind smaller gaps in sensibility that depend on $m_\phi$ and the couplings. For $\frac{\muu^2 L}{4\omega}\approx \pi n\gg 1$, the gaps in sensitivity can also be removed by a small displacement of the detector or, as we already mentioned, using data with the magnetic field switched on (subleading effects become relevant). The case when $\frac{\muu^2 L}{4\omega}\approx \pi n$, with $n$ a small integer, is more complicated because the switching on of the magnetic field may still leave gaps and in this case moving the detector a small distance would have no mayor impact. This means that the most effective way of filling up the sensitivity gaps for $n$ a small integer is by using a gas that changes the refraction index in the chamber.

\begin{figure}
\center
\begin{subfigure}{.9\textwidth}
\adjustbox{trim={.01\width} {.005\height} {0\width} {.015\height},clip}{\includegraphics[width=1\linewidth]{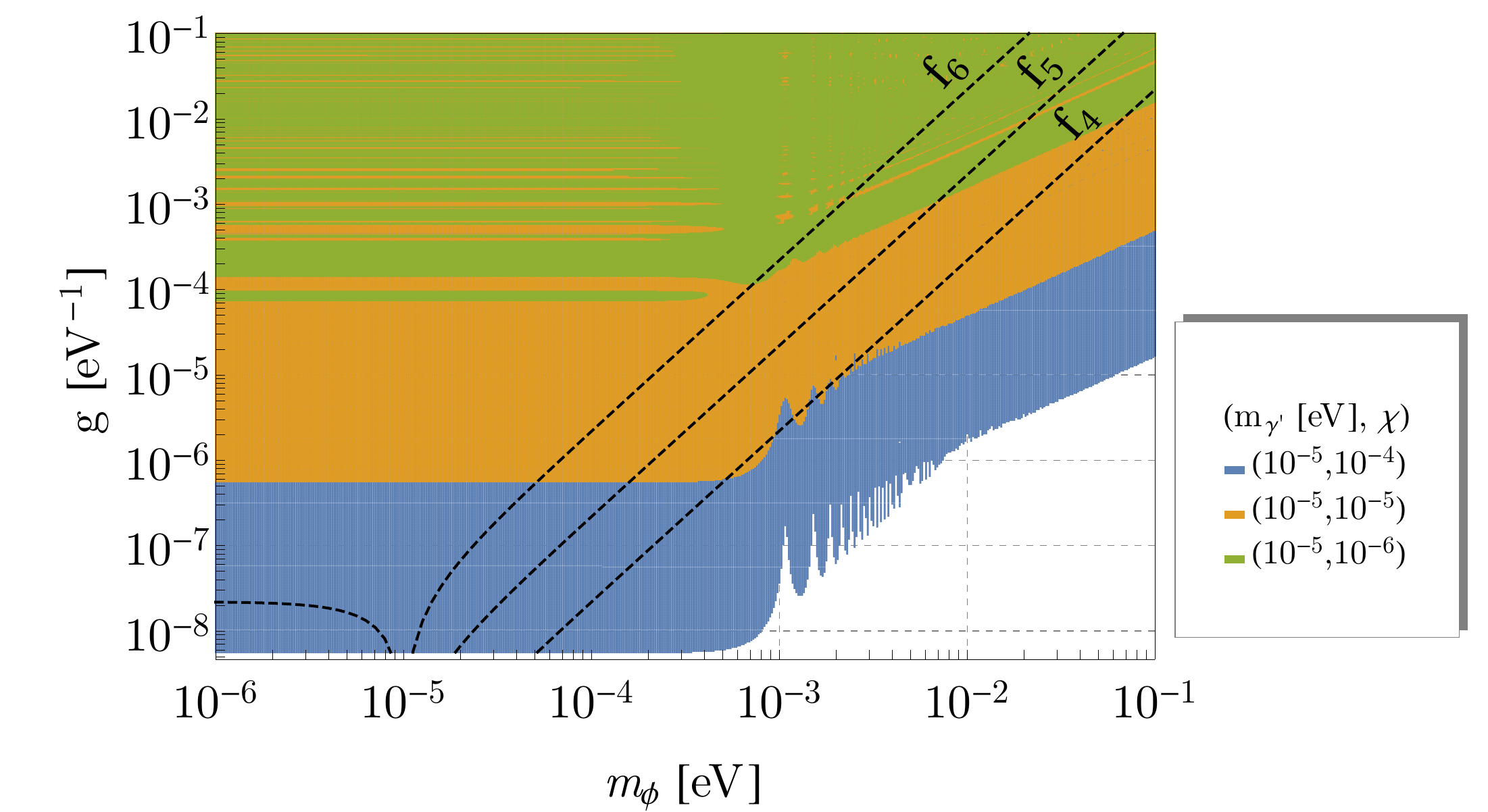}}
\end{subfigure}
\begin{subfigure}{.9\textwidth}
\adjustbox{trim={.01\width} {.005\height} {0\width} {.015\height},clip}{\includegraphics[width=1\linewidth]{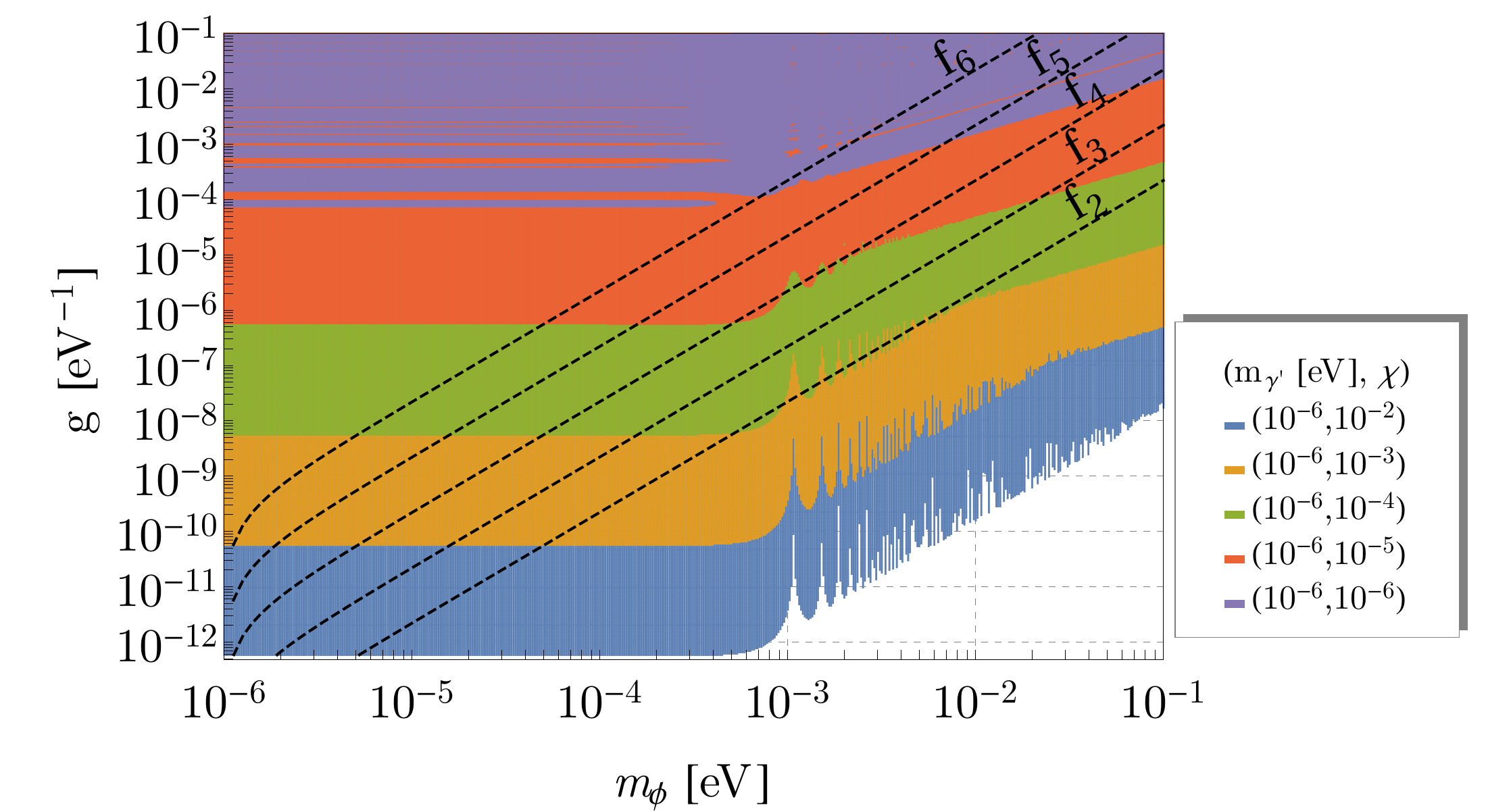}}
\end{subfigure}
\caption{\small{Exclusion plots for axion-like particles using a LSW setup with parameters $L=5.5$~m, $B=5.5$~T, $\omega=2.33$~eV. We considered HP masses $\muu=10^{-5},10^{-6} $~eV and several coupling to photons. The dashed lines corresponds to $f_i\rightarrow|x|=\chi=10^{-i}$. Over the dashed line we have $|x|<\chi$, thus a strong mixing regime and under the dashed line we have the weak mixing regime. The empty stripes and empty rounded zones correspond to sensitivity gaps.}}
\label{fig:alps1,(mphi,g)-plane}
\end{figure}

\begin{figure}
\center
\begin{subfigure}{.9\textwidth}
\adjustbox{trim={.02\width} {.005\height} {0\width} {.015\height},clip}{\includegraphics[width=1\linewidth]{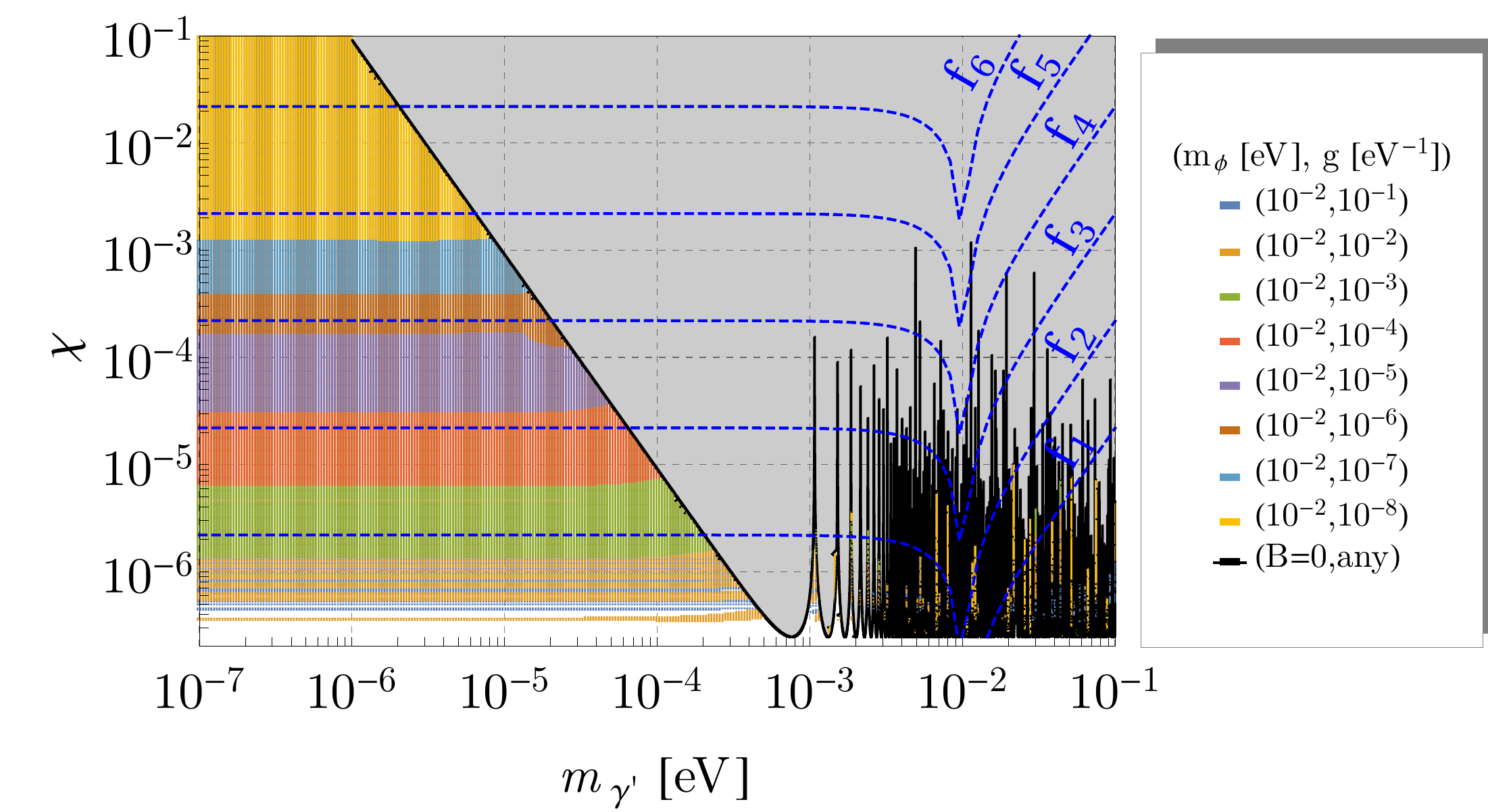}}
\end{subfigure}
\begin{subfigure}{.9\textwidth}
\adjustbox{trim={.02\width} {.005\height} {0\width} {.015\height},clip}{\includegraphics[width=1\linewidth]{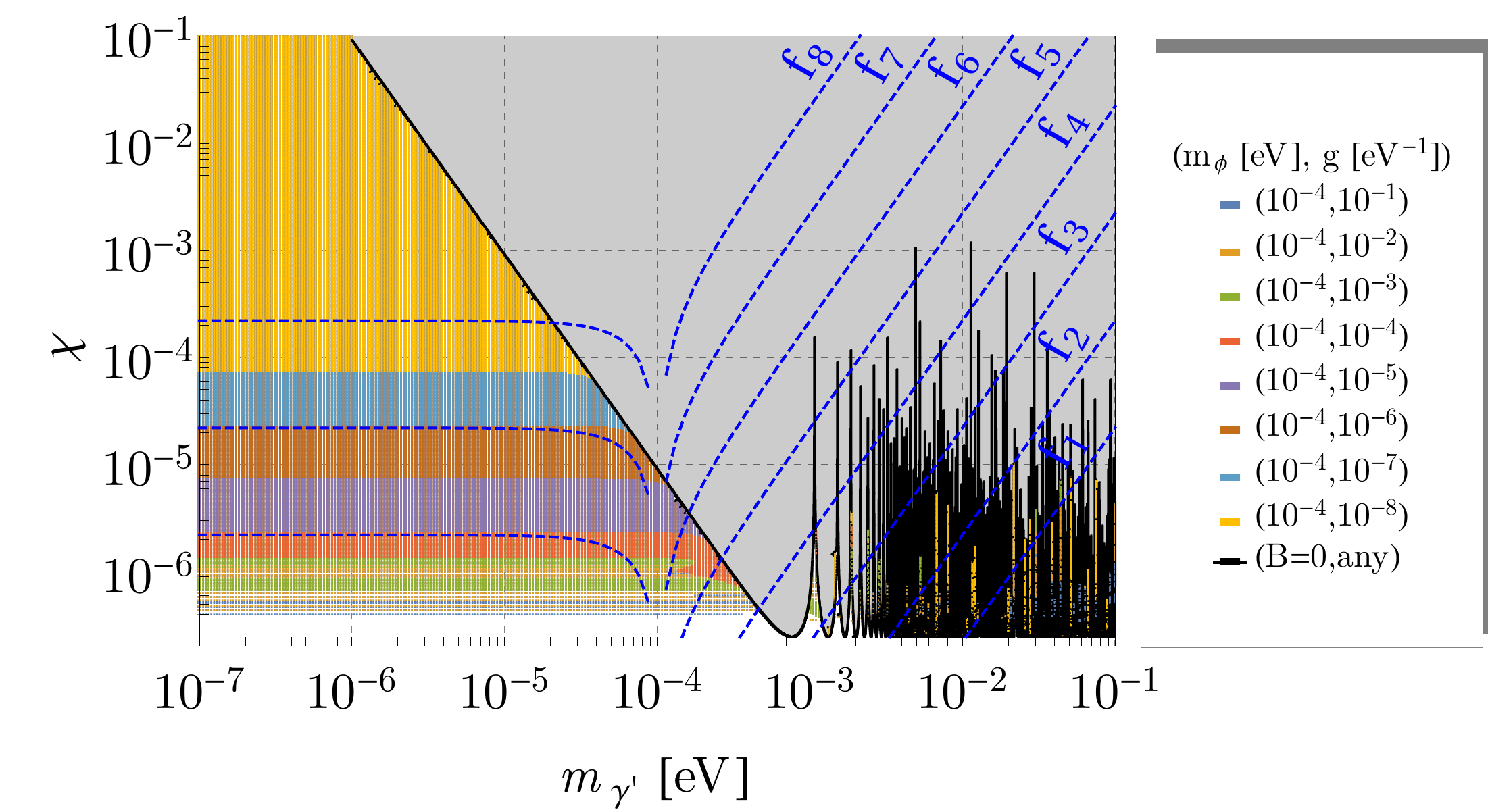}}
\end{subfigure}
\caption{Excluded regions for different masses of the ALP and the ALP-coupling constant using ALPS-I results. The black line and shaded upper region is the exclusion zone for zero magnetic field. In this case the axion of our model decouples and so this excluded region extends to any value of $m_\phi$ and $g$. The dashed lines corresponds to $f_i\rightarrow|x|=\chi=10^{-i}$. For masses $m_\phi\lesssim 10^{-4}$ eV, all the excluded regions are the same as for the massless case.}
\label{fig:alps1,(mx,chi)-plane}
\end{figure}

\subsection{Bounds from experiments}

In figure~(\ref{fig:alps1,(mphi,g)-plane}) we have plotted the sensitivity for the ALP parameter space, using typical values of a first-generation LSW experiment (see details in the figures). The dashed lines corresponds to $|x|=\chi$, and over (under) corresponds to strong (weak) mixing regime or $|x|<\chi$ ($|x|>\chi$). Over the dashed line, in the strong mixing regime, we can see gaps in sensitivity that correspond to horizontal lines. The plots confirm that, in the weak mixing regime (under the dashed line), the oscillations start to dominate the probability when $m_\phi^2 L/(4\omega) >1$ and the expected the sensitivity to the ALP gets poorer.

In order to obtain bounds for the model presented here, we used the results of the experiment ALPS-I. In \cite{ALPS-I}, the authors report measurements of photon counting in the second region of a LSW-type of experiment with and without the magnetic field turned on. For the run with $B=0$, the bounds they can impose on the HP will also apply to this model, see shaded region in plots of figure (\ref{fig:alps1,(mx,chi)-plane}) (in this case the plot fig. 5 in \cite{ALPS-I}), but the bounds obtained by  runs with magnetic field on have to be recalculated. Our results are summarised in figures (\ref{fig:alps1,(mphi,g)-plane}) and (\ref{fig:alps1,(mx,chi)-plane}).

For small couplings, the region $|x|\sim \chi$ (dashed line in plots of figures (\ref{fig:alps1,(mphi,g)-plane}) and (\ref{fig:alps1,(mx,chi)-plane})), give us a rough estimate of the limits in sensitivity of the experimental setup of a LSW-type experiment. 

For the hidden photon, the bounds in this model seem quite tight, considering they are obtained only using optical laboratory experiments in vacuum. The  low mass region of the HP gets fully covered for $\chi\lesssim 10^{-6}$, thanks to the oscillation of HP-photon induced by the magnetic field. It is therefore interesting to find out if more constraining observations, such as solar evolution, indeed rule out also an  important fraction of  the low mass region (besides the expected keV region). For the ALP case, the constraints on $g$ for small masses regions go as low as $g\sim 10^{-14}$~$eV^{-1}$ when the corresponding HP as a coupling constant of $\chi \lesssim 10^{-1}$. Therefore, the effective coupling to photons that can be probed, given by $\chi^2 g$ is of the order of $\chi^2 g\lesssim10^{-7}$~GeV$^{-1}$ for the small mass region. This constraint is quite similar to the one that was imposed for the model ALP-photon alone using LSW experiments. The reason seems to be the resonant effect that can appear at low masses, where both ALP and HP are strongly coupled. For bigger masses the oscillations in the mass start to decrease the sensitivity to ALPs. For the $\mfi\gtrsim 10^{-3}~$eV the constraints on $g\chi^2$ are a bit worst than the ones on the ALP-photon model. 

For the birefringence effects, we have used the latest results of the PVLAS experiment \cite{pvlas}. We find that the constraints on the model are always less stringent than the LSW ones. Nonetheless, we show the constraints on the ALP for different masses and coupling of the HP, for completeness, see figures (\ref{fig:pvlas1}) and (\ref{fig:pvlas2})
\section{Conclusions}{\label{conclu}}

We have presented a detailed analysis on a model of coupled HP and ALP particles and their mixing with photons in the presence of a magnetic field in vacuum. We have found an interesting three particle mixing scenario, with two oscillation angles, the usual $\chi$ that parametrises the strength of the kinetic mixing between HP and photons and a second mixing angle, given by eq.~(\ref{theta2}), which controls the oscillation between the ALP and the linear combination of HP and photon $\cos\chi X_\parallel-\sin\chi A_\parallel$. This angle can be maximised in the presence of very strong magnetic fields (and small masses $\mfi,\,\muu \lesssim 10^{-5}~$eV), and the probability of conversion of the photon, either to ALP or HP gets its  highest value in vacuum. It seemed instructive to start with the simple vacuum interaction to get a sense of the most important features, since they more or less reappear in when considering a medium, as for example, an electron plasma.  It is of course expected that if media is present, new resonances could emerge, and hopefully, the model could be extensively constrained in parameter space. 

\begin{figure}[ht!]
\begin{subfigure}{.5\textwidth}
\adjustbox{trim={.005\width} {0\height} {0\width} {0\height},clip}{\includegraphics[width=1\linewidth]{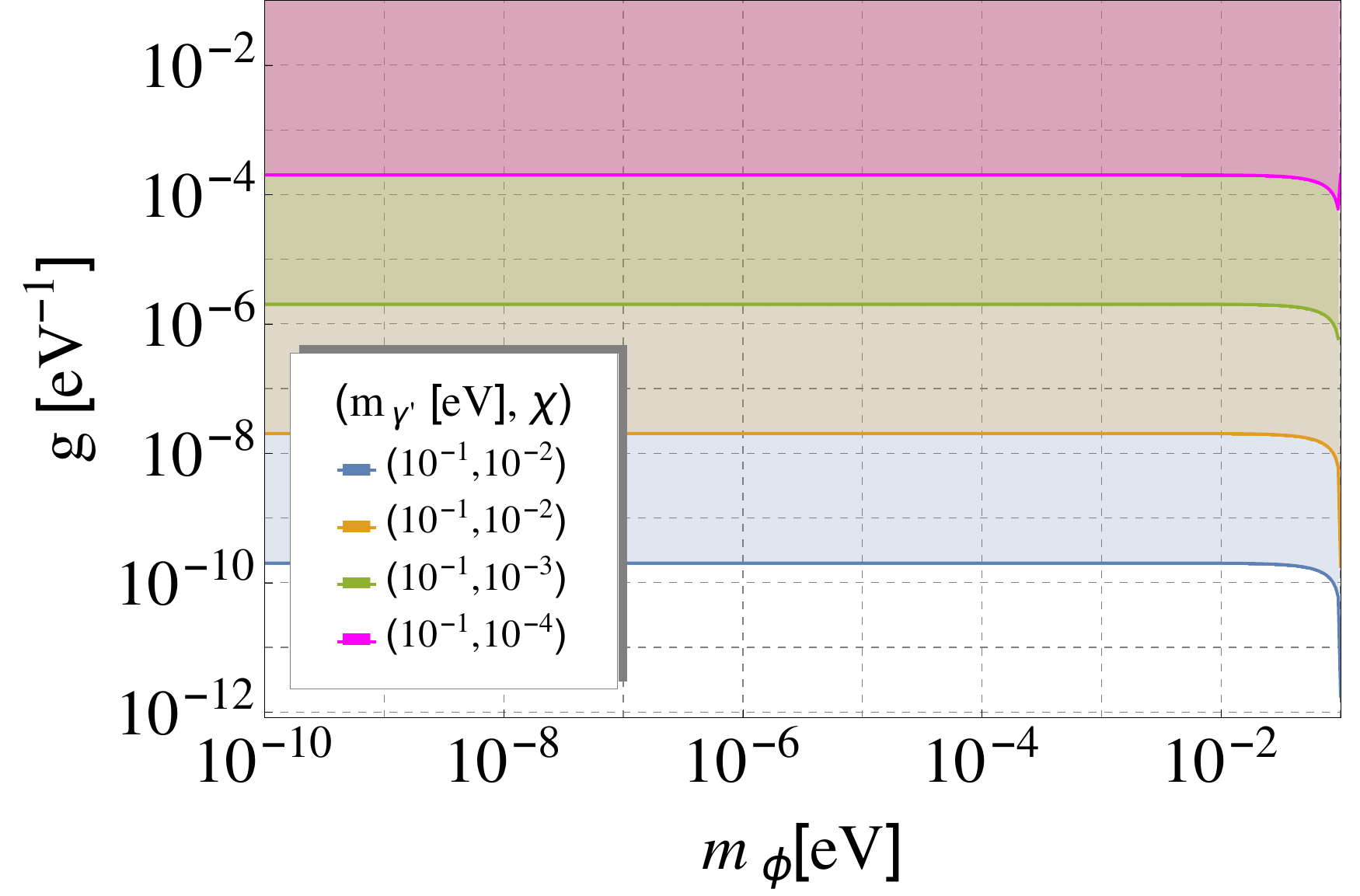}}
\end{subfigure}
\begin{subfigure}{.5\textwidth}
\adjustbox{trim={.005\width} {0\height} {0\width} {0\height},clip}{\includegraphics[width=1\linewidth]{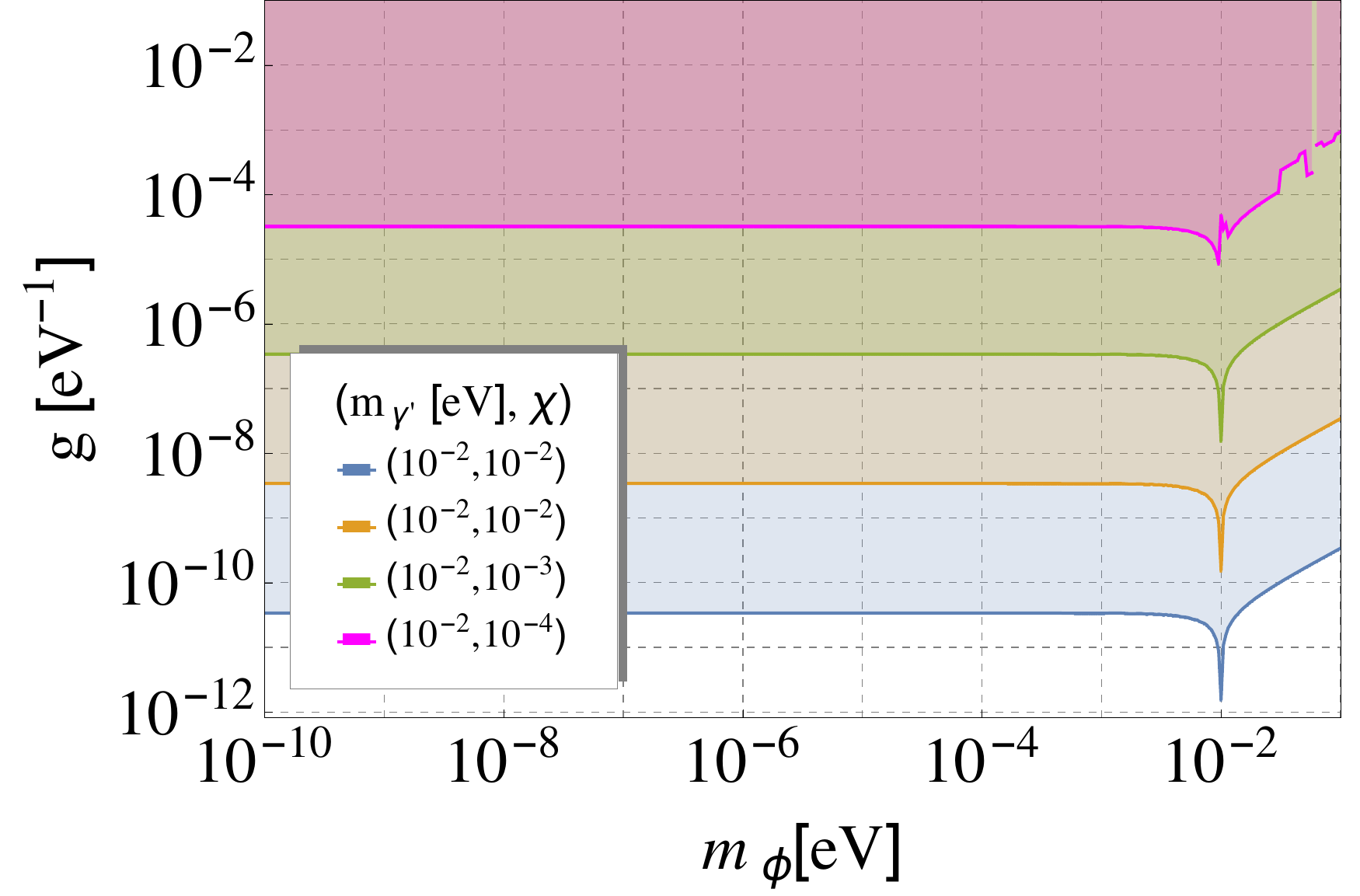}}
\end{subfigure}
\begin{subfigure}{.5\textwidth}
\adjustbox{trim={.005\width} {.005\height} {0\width} {0\height},clip}{\includegraphics[width=1\linewidth]{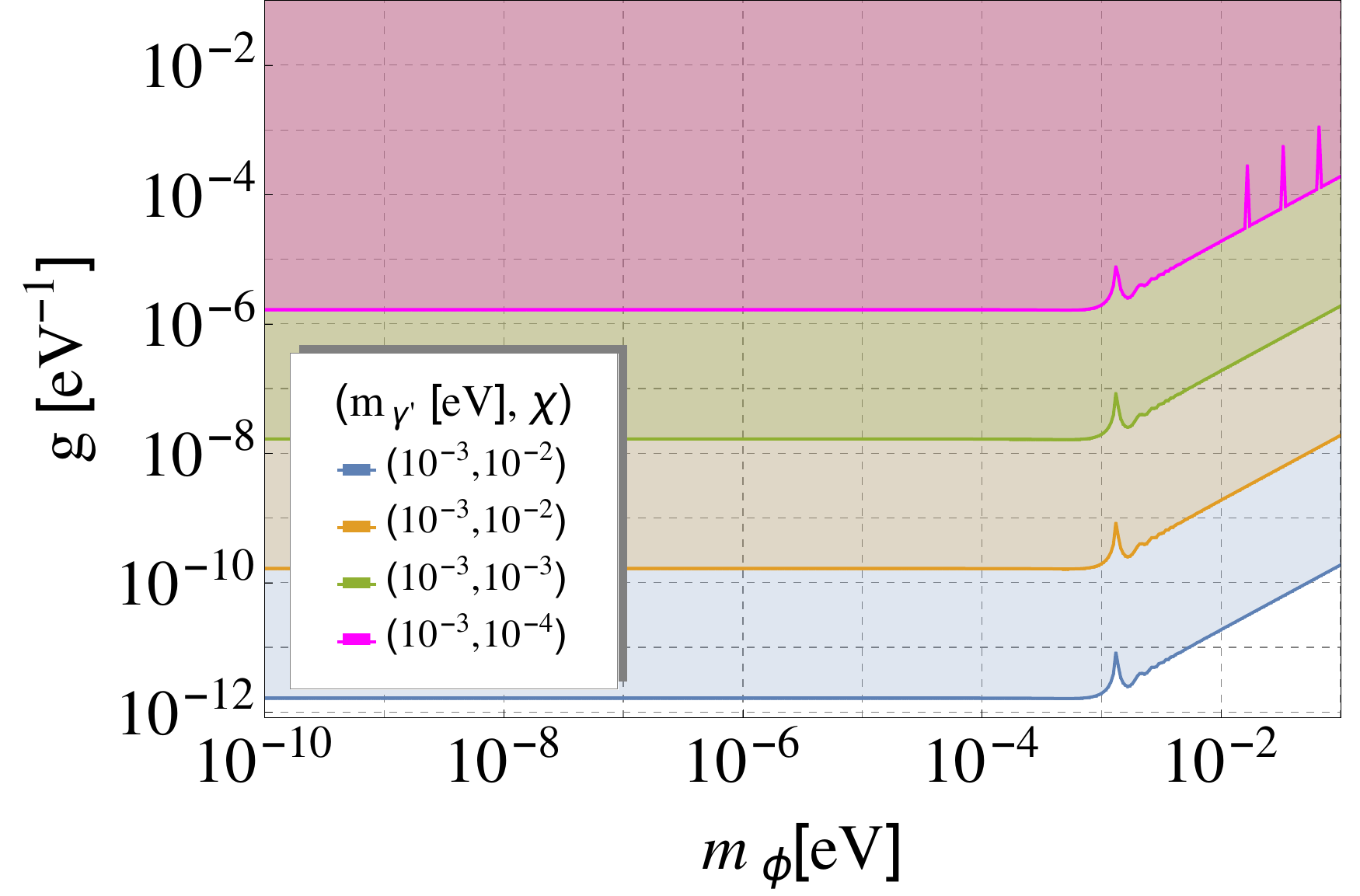}}
\end{subfigure}
\begin{subfigure}{.5\textwidth}
\adjustbox{trim={.005\width} {.005\height} {0\width} {0\height},clip}{\includegraphics[width=1\linewidth]{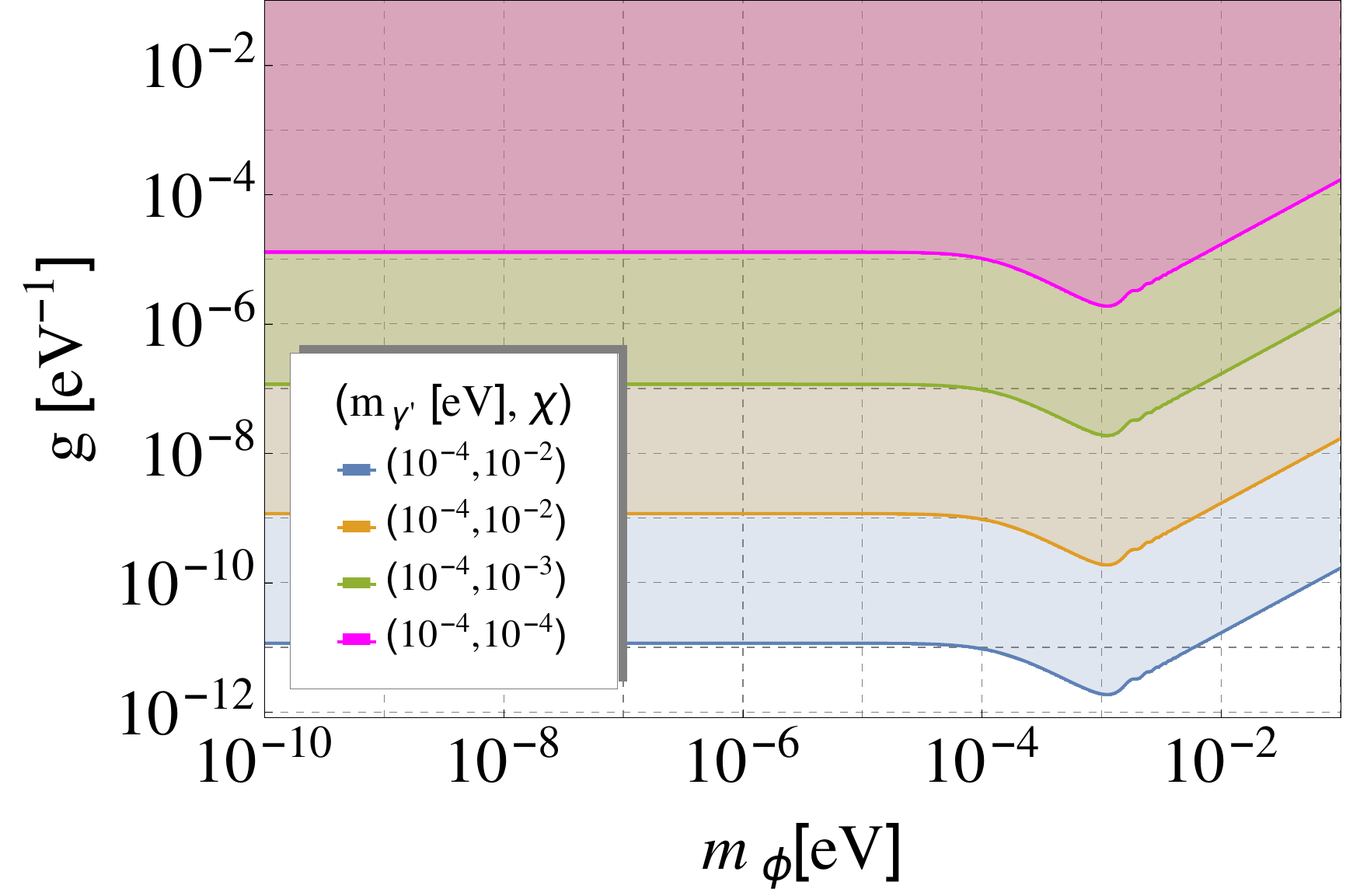}}
\end{subfigure}
\caption{Exclusion region in the ALP plane region $g$ vs. $\mfi$ using ellipticity measurements for different pairs of HP coupling and masses. We have assumed $B=2.5~$T, $L=1$~m and $\omega=1$~eV from \cite{pvlas}. See discussion in subsection 4.1 of the manuscript. }
\label{fig:pvlas1}
\end{figure}

Let us also mention that a main motivation to introduce this was to explain the 3.55 keV line reported in 2014 \cite{Jaeckel:2014qea}. There, it was assumed the ALP is the CDM and the HP could be in principle massless. Thus, another interesting scenario to test this model is precisely to compute the relic abundance of both ALP and HP in the early universe and the evolution of the fields. \\
\vskip 2mm

\noindent Acknowledgments: This work has been supported by FONDECYT project 1161150.

\begin{figure}[ht!]
\begin{subfigure}{.5\textwidth}
\adjustbox{trim={.005\width} {.005\height} {0\width} {0\height},clip}{\includegraphics[width=1\linewidth]{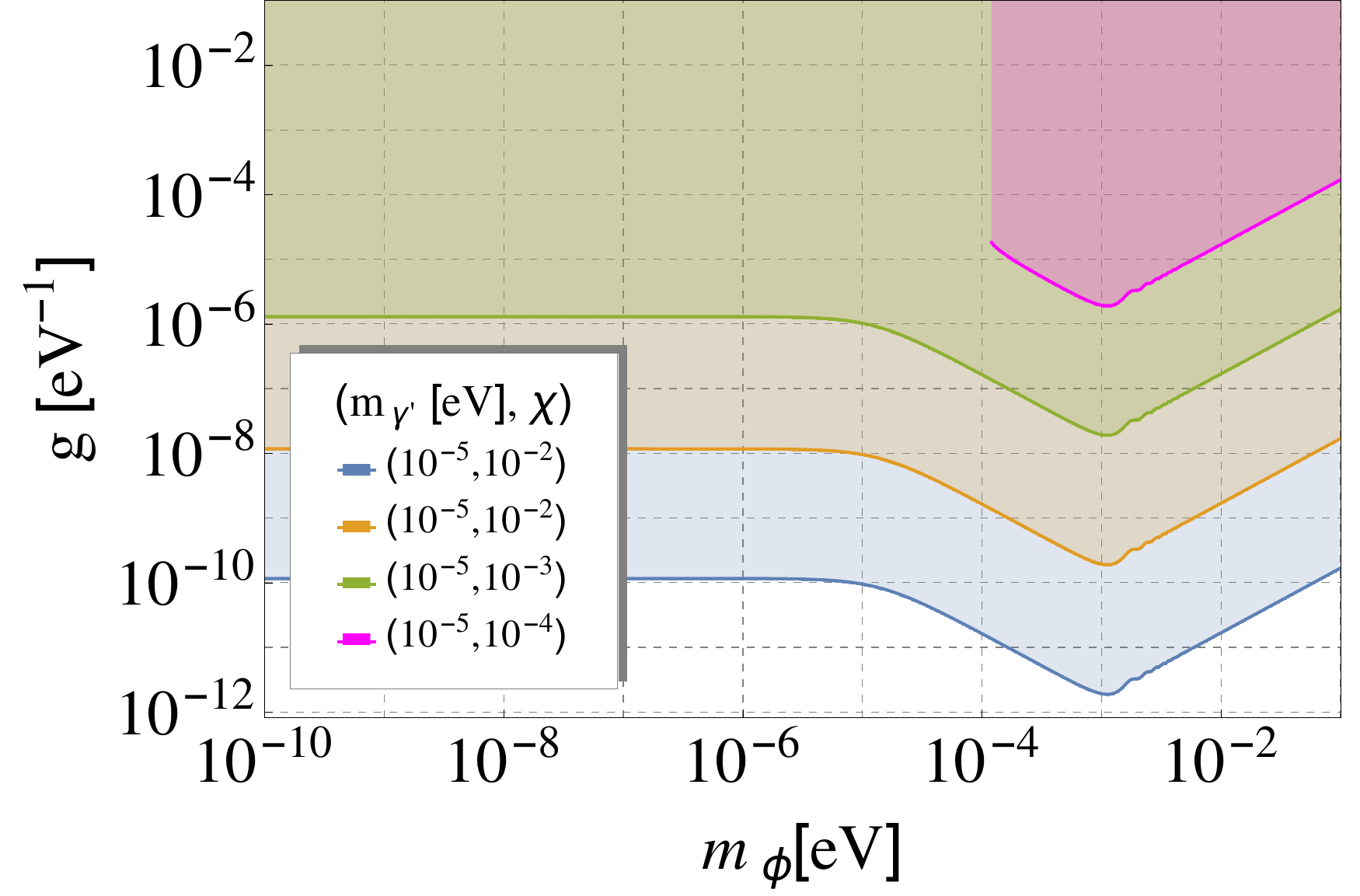}}
\end{subfigure}
\begin{subfigure}{.5\textwidth}
\adjustbox{trim={.005\width} {.005\height} {0\width} {0\height},clip}{\includegraphics[width=1\linewidth]{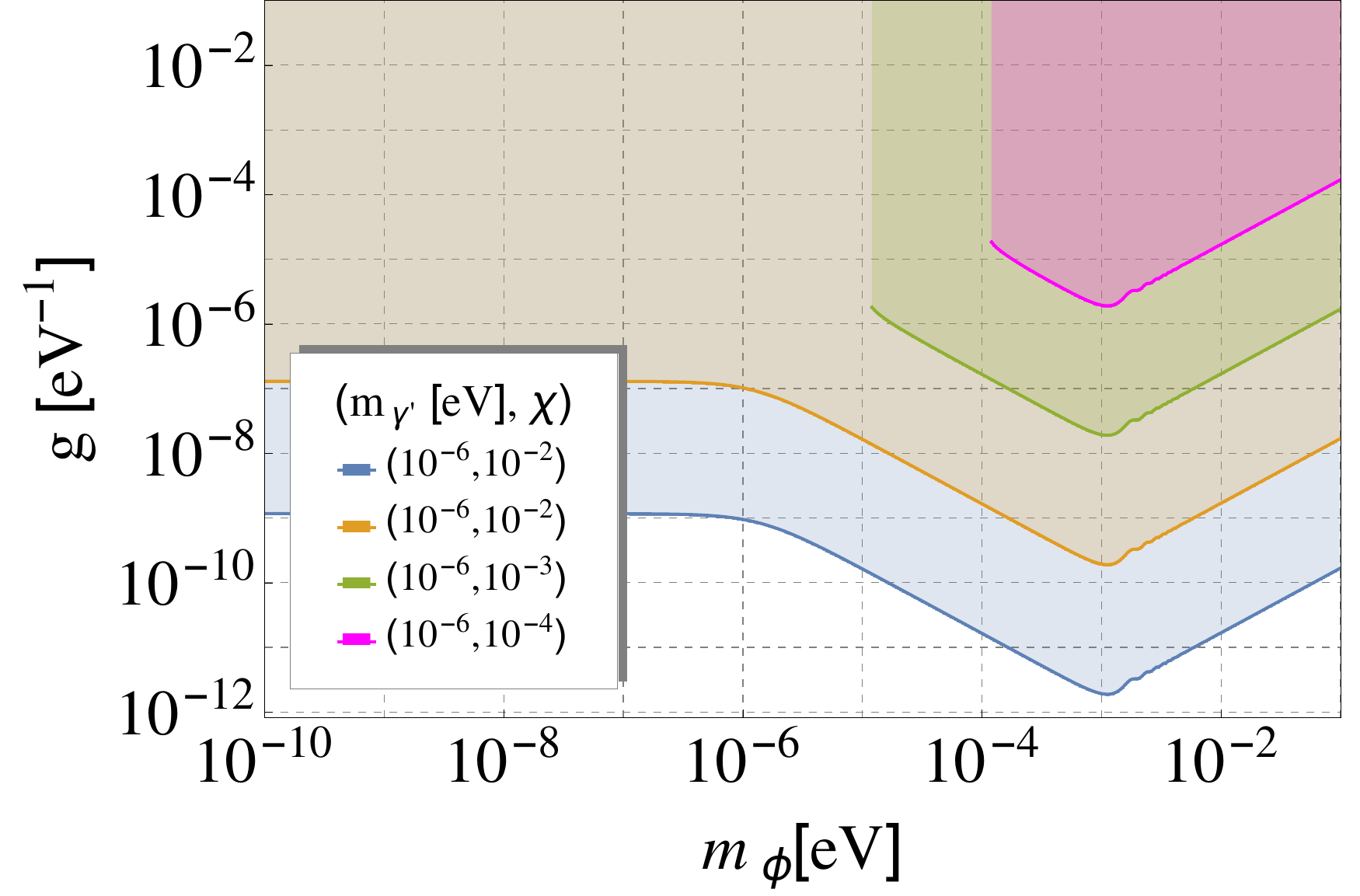}}
\end{subfigure}
\begin{subfigure}{.5\textwidth}
\adjustbox{trim={.005\width} {.005\height} {0\width} {0\height},clip}{\includegraphics[width=1\linewidth]{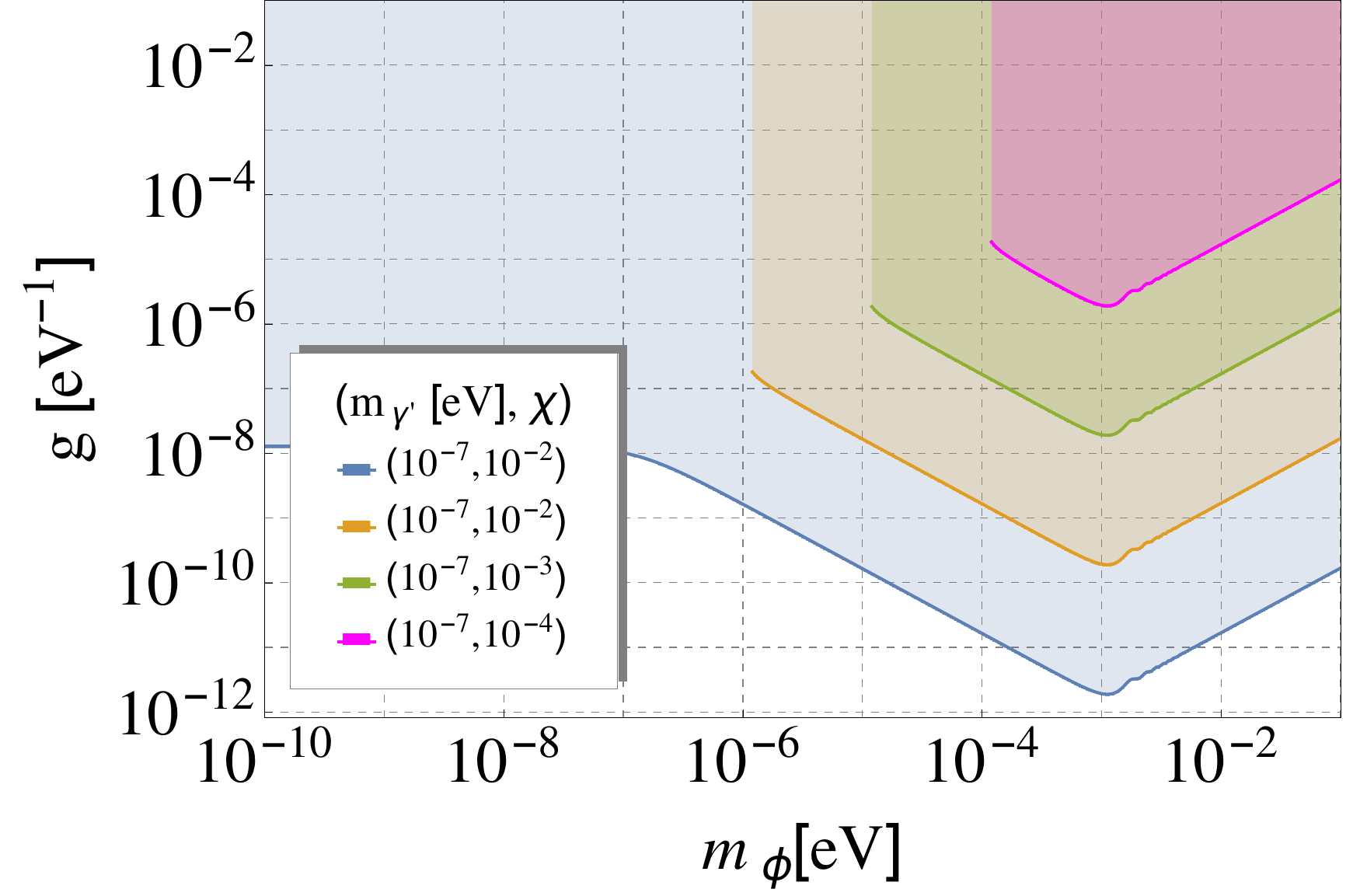}}
\end{subfigure}
\begin{subfigure}{.5\textwidth}
\adjustbox{trim={.005\width} {.005\height} {0\width} {0\height},clip}{\includegraphics[width=1\linewidth]{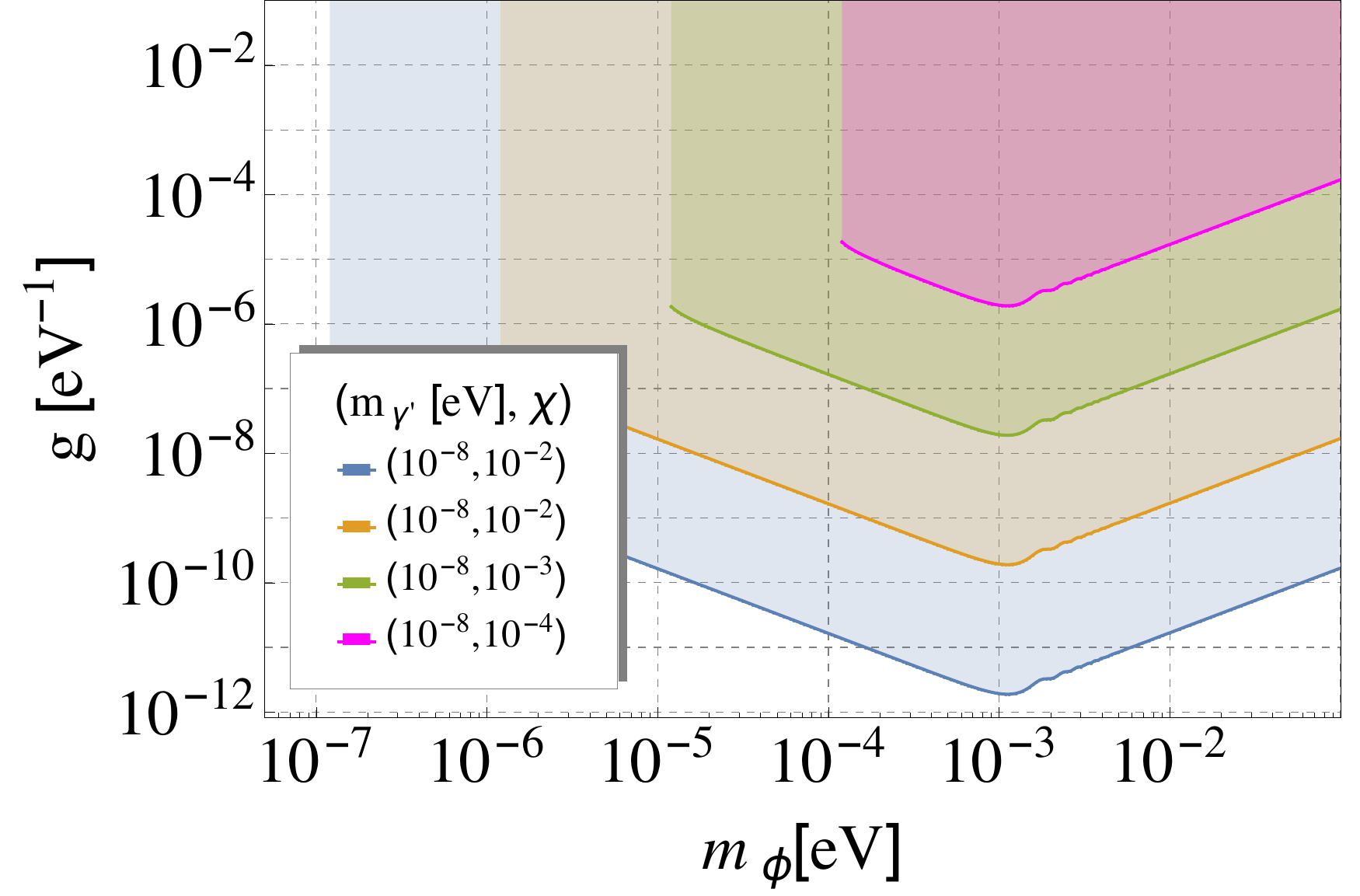}}
\end{subfigure}
\begin{subfigure}{.5\textwidth}
\adjustbox{trim={.005\width} {.005\height} {0\width} {0\height},clip}{\includegraphics[width=1\linewidth]{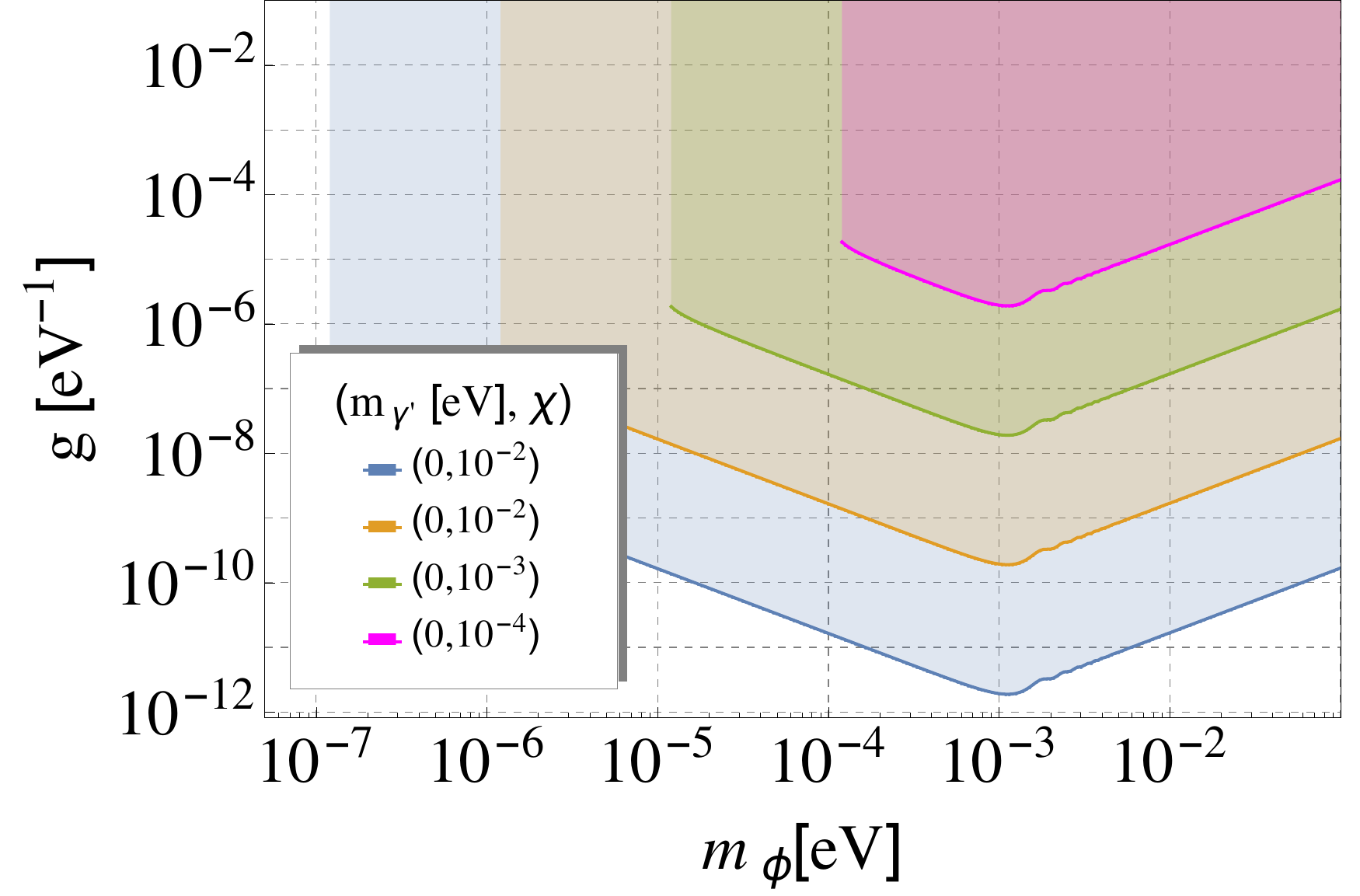}}
\end{subfigure}
\caption{Exclusion region in the ALP plane region $g$ vs. $\mfi$ using ellipticity measurements for different pairs of HP coupling and masses. We have assumed $B=2.5~$T, $L=1$~m and $\omega=1$~eV from \cite{pvlas}. See discussion in subsection 4.1 of the manuscript.}
\label{fig:pvlas2}
\end{figure}


\end{document}